\documentclass[journal]{IEEEtran}

\usepackage{booktabs}
\usepackage{graphicx}
\graphicspath{{fig/}}
\usepackage{cite}
\usepackage{makecell}
\usepackage{amsmath,amssymb,amsfonts}
\usepackage{algorithmic}
\usepackage{textcomp}
\usepackage{subfigure}
\usepackage[ruled]{algorithm2e}
\usepackage{multirow}
\usepackage{tabularx}
\usepackage{bm}
\usepackage{soul}
\usepackage{balance}
\soulregister\cite7
\soulregister\citep7
\soulregister\citet7
\soulregister\ref7
\soulregister\pageref7

\usepackage{xcolor,soul,framed} 
\colorlet{shadecolor}{yellow}

\begin{document}

	\title{ Improved Soft-$k$-Means Clustering Algorithm for Balancing Energy Consumption in Wireless Sensor Networks}

\author{Botao~Zhu,~\IEEEmembership{}
      Ebrahim~Bedeer,~\IEEEmembership{Member,~IEEE,}
      ~Ha H. Nguyen,~\IEEEmembership{Senior Member,~IEEE,} \\Robert Barton,~\IEEEmembership{}and Jerome Henry~\IEEEmembership{}
\thanks{B. Zhu, E. Bedeer, and H. H. Nguyen are with the Department of Electrical and Computer Engineering, University of Saskatchewan, Saskatoon, Canada S7N 5A9. Emails: \{botao.zhu, e.bedeer, ha.nguyen\}@usask.ca}
\thanks{R. Barton and J. Henry are with Cisco Systems Inc. Emails: \{robbarto, jerhenry\}@cisco.com.}
\thanks{This work was supported by NSERC/Cisco Industrial Research Chair in Low-Power Wireless Access for Sensor Networks.}
\thanks{Copyright (c) 20xx IEEE. Personal use of this material is permitted. However, permission to use this material for any other purposes must be obtained from the IEEE by sending a request to pubs-permissions@ieee.org.}
}

\maketitle
\begin{abstract}
	 Energy load balancing is an essential issue in designing wireless sensor networks (WSNs). Clustering techniques are utilized as energy-efficient methods to balance the network energy and prolong its lifetime. In this paper, we propose an improved soft-$\it{k}$-means (IS-$\it{k}$-means) clustering algorithm to balance the energy consumption of nodes in WSNs. First, we use the idea of ``clustering by fast search and find of density peaks'' (CFSFDP) and kernel density estimation (KDE) to improve the selection of the initial cluster centers of the soft $\it{k}$-means clustering algorithm. Then, we utilize the flexibility of the soft-$\it{k}$-means and reassign member nodes considering their membership probabilities at the boundary of clusters to balance the number of nodes per cluster. Furthermore, the concept of multi-cluster heads is employed to balance the energy consumption within clusters. {Extensive simulation results under different network scenarios demonstrate that for small-scale WSNs with single-hop transmission}, the proposed algorithm can postpone the first node death, the half of nodes death, and the last node death on average when compared to various clustering algorithms from the literature.
\end{abstract}

\begin{IEEEkeywords}
	Clustering by fast search and find of density peaks (CFSFDP), energy load balancing, kernel density estimation (KDE), multi-cluster heads, soft $\it{k}$-means, wireless sensor networks (WSNs).
\end{IEEEkeywords}

\section{Introduction}
\label{sec:introduction}
\IEEEPARstart{T}{he} {general concept of Internet-of-Things (IoT) is to facilitate the network connection of billions of devices to collect and exchange information to provide various services \cite{L. Chettri, M. Stoyanova}. Wireless sensor networks (WSNs) are among important parts of an IoT system because they can be used to gather and send data \cite{O. Elijah}. WSNs are like the eyes and ears of the IoT and they build the bridge between the real and the digital worlds. WSNs typically consist of a large number of low-cost sensor nodes with restricted battery supplies. Sensor nodes are deployed in various application scenarios to monitor and collect physical conditions of the surrounding environment such as temperature, humidity, pressure, position, vibration, and sound,  to name a few \cite{S. K. Mohapatra}. The collected data is then sent to the base station (BS) for further analysis and processing}.

Reducing energy consumption is a key challenge in WSNs, as sensor nodes can be placed in hard-to-reach areas and/or their batteries may not be rechargeable \cite{b1}. Clustering in energy-limited WSNs has been widely investigated to reduce the energy consumption \cite{b22}. Clustering-based algorithms group sensor nodes into distinct clusters, where each sensor node belongs to one cluster only. All member nodes sense their surrounding environment and send the results to the cluster heads (CHs). Then, CHs collect and process the data and send information to the BS\cite{b2}. Each node consumes a certain amount of energy when it collects, processes, and sends data, and a node is defined to be dead when it runs out of energy \cite{b3}. Hence, it is crucial to develop efficient clustering algorithms to balance the energy consumption among sensor nodes in WSNs.

Different clustering techniques have been proposed to design energy-efficient WSNs and increase their lifetime. {The authors in \cite{T. M. Behera} proposed a CH election method, which rotates the CH positions among the nodes with higher energy in different communication rounds. In particular, the method considers the initial energy, residual energy, and an optimal number of CHs to decide the next group of CHs among the nodes in the network. Then, member nodes join different CHs according to the distances between them and CHs to form clusters. A joint clustering and routing algorithm is proposed in \cite{Z. Xu} to improve the energy efficiency of large-scale WSNs. This algorithm employs a back-off timer and gradient routing to execute the CH selection and multi-hop routing simultaneously. The authors in \cite{R. Zhang J. Pan} presented a node-density-based clustering and mobile elements algorithm (NDCMC) for collecting data in WSNs. In NDCMC, the nodes surrounded by more deployed nodes are selected as CHs in order to improve the efficiency of intra-cluster routing. The authors presented a fixed parameter tractable (FPT) approximation algorithm with an approximation factor of 1.2 based on the parameterized complexity theory in \cite{R. Yarinezhad} in order to solve load balanced clustering problem (LBCP) in WSNs. The FPT-approximation algorithm determines which gateway each sensor node must be assigned to, which can lead to more balanced load and energy consumption among the gateways. On the other hand, a routing tree for the inter-cluster communication is proposed, which can distribute the overhead of the routing among almost of all of the nodes. In \cite{R. Yarinezhad and S. N. Hashemi}, the authors further proposed an FPT-approximation algorithm with an approximation factor of 1.1, which is more precise than previous approximation factors reported for LBCP. The FPT-approximation algorithm is used to assign sensor nodes to gateways such that the maximum load of the gateways is minimized. Then, an energy-aware routing algorithm is employed to find the optimal routing tree between gateways and the sink with the aim of balancing the energy consumption of the nodes. The same authors also considered another FPT-approximation algorithm with an approximation factor of 1.1 in \cite{S. N. Hashemi}. In order to make the FPT-approximation algorithm to be practical in large-scale WSNs, a virtual grid infrastructure with several equal-size cells is used where the FPT-approximation algorithm runs in each cell independently. In \cite{N. Mazumdar}, a distributed multi-objective based clustering algorithm is presented to assign sensor nodes to appropriate CHs. Then, an energy-efficient routing algorithm is proposed to balance the relay load among the CHs. In \cite{H. Om}, the authors implemented a distributed clustering algorithm by considering a trade-off between the energy efficiency and coverage requirement. This algorithm can form unequal-size clusters to balance the load of the CHs.  The same authors in \cite{N. Mazumdar and H. Om} proposed a distributed fuzzy logic-based unequal clustering approach and routing algorithm (DFCR) to solve the hot spot problem, which is caused by the fact that some CHs deplete their energy much faster as compared to other CHs. The DFCR algorithm designs an unequal clustering mechanism by reducing the cluster size nearest to the BS.}

{The authors in  \cite{b12} proposed a modified $k$-means clustering algorithm that considers two factors, namely, (i) distances among CHs and their member nodes, and (ii) the remaining energy of nodes, to reduce the overall energy consumption and extend the network lifespan. In  \cite{b14}, the authors proposed a hybrid clustering algorithm based on the $k$-means clustering algorithm and LEACH\cite{b10}, where balanced clusters are generated by $k$-means and CHs are selected by LEACH. This hybrid algorithm outperforms LEACH in terms of the energy consumption. However, due to the frequent re-clustering, the energy consumption of the nodes may increase in the phase of cluster formation and CH selection. An energy efficient clustering protocol based on $k$-means (EECPK-means) is proposed in  \cite{b15} with the aim of  balancing the load of CHs in WSNs. The midpoint method is used to improve the initial selection of centroids in the $k$-means algorithm in order to generate balanced clusters. In \cite{N. T. Tam}, the authors proposed a method based on fuzzy $c$-means clustering and particle swarm optimization (FCM-PSO) to reduce the total energy consumption of the network and reduce the number of network disconnects. The FCM-PSO algorithm considers the energy consumption and constraints of communication in the calculations of the CHs and nodes’ membership probability.} The energy-efficient $k$-means LEACH (KM-LEACH) algorithm is proposed in \cite{b5} to create symmetric clusters and reduce the average intra-cluster communication distance, which can save nodes' energy and improve the network lifetime. {To address the problem of how to control the failure of a CH in each cluster}, the $k$-medoids clustering algorithm and vice CH scheme (VLEACH) are used together with LEACH in \cite{b6}. Vice CH will become a new CH in case the CH of a given cluster dies, which helps to prolong the lifetime of WSNs by balancing the nodes' energy consumption. The authors in  \cite{b7} used the $k$-means and Gaussian elimination algorithms to reduce energy consumption of WSNs and extend their lifetime. An innovative classification algorithm based on
``clustering by fast search and finding of density peaks" (CFSFDP)\cite{b18} algorithm for balancing energy is proposed in \cite{Y. Zhang}. The authors extend the original CFSFDP algorithm to take into account residual energy (in addition to local density and distance) to select CHs, and accordingly cluster nodes based on the selected CHs.

Against the above background, in this paper, an improved soft-$k$-means (IS-$k$-means) clustering algorithm is proposed {with the aim of balancing the energy consumption of all nodes in WSNs and extending the network lifetime. The proposed IS-$k$-means can be widely used in industrial control, smart home, smart agriculture, environment perception, health monitoring, etc., because it can extend the life of sensor nodes in these application scenarios.} The novelty of the proposed algorithm can be summarized as follows.

1) Compared with existing clustering algorithms that select the initial cluster centers randomly, we choose the initial centroids of the IS-$k$-means clustering algorithm by using the idea of density from CFSFDP and kernel density estimation (KDE) \cite{b29} to achieve a better clustering result. The nodes with high local density and relative large node distances are chosen as the initial centroids.

2) After the proposed algorithm converges, we reassign member nodes that are located at the boundary of two or more clusters to balance the number of nodes per cluster according to the flexibility of the soft-$\it{k}$-means.

3) Since the clustering process needs to be repeated continually, the communication cost during the clustering phase is increased. We use multi-cluster heads (multi-CHs) scheme to balance traffic load of CHs of different clusters and reduce the frequency of clustering.

The rest of this paper is organized as follows. The necessary background for our research is discussed in Section \uppercase\expandafter{\romannumeral2}. Section \uppercase\expandafter{\romannumeral3} describes the proposed IS-$k$-means algorithm. In Section \uppercase\expandafter{\romannumeral4}, we compare the performance of the proposed IS-$k$-means with other algorithms. Finally, Section \uppercase\expandafter{\romannumeral5} concludes the paper.

\section{Preliminaries}

\subsection{Soft $k$-Means}

The soft $k$-means \cite{b8} is a kind of fuzzy clustering algorithm where clusters are represented by their respective centers. Since traditional $k$-means clustering techniques are hard clustering algorithms, which may fail to separate overlapping clusters or properly cluster noisy data \cite{b24}, the soft $k$-means algorithm can be applied to address these cases. With the soft $k$-means algorithm, each node may belong to one or more clusters with different degrees of membership \cite{b25}. Nodes located at the boundaries of clusters are not forced to fully belong to a given cluster, but rather they can be members of many clusters with membership degrees or probabilities between 0 and 1 \cite{b23}. Nodes at the edge of a cluster may have lower membership probabilities than nodes close to the center of a cluster. This flexibility of the soft $k$-means clustering is in sharp contrast with the $k$-means clustering, where a node belongs to only a single cluster.

For a set of nodes' locations $\bm{X} = \{\bm{x}_1,\bm{x}_2,\dots,\bm{x}_n \}$ in WSNs, the goal of the soft $k$-means is to partition the $n$ nodes into $k$ sets $\bm{C} = \{\bm{c}_1,\bm{c}_2,\dots,\bm{c}_k\}$ with small intra-cluster distances and large inter-cluster distances. Thus, we define the following cost function:
\begin{equation}
\label{eq_object}
J(\bm{X};\bm{Z},\bm{M}) = \sum_{v=1}^{k}\sum_{j=1}^{n}z_{vj} ||\bm{x}_j-\bm{\mu}_{v}||^2,
\end{equation}
where $\bm{M}(\bm{\mu}_{v};{v}=1,\dots,k)$ is the matrix of cluster centers, and $\bm{Z}(z_{{v}j};{v}=1,\dots,k;j=1,\dots,n)$ is the membership probability matrix of $\bm{X}$. $z_{vj}$ is the membership value of the $j$th node to the $v$th cluster and is defined as  \cite{b8}
\begin{equation}
\label{eq1}
z_{vj} = \frac{e^{-\beta ||\bm{x}_{j}-\bm{\mu}_{v}||^2}}{\sum_{l=1}^{k}e^{-\beta||\bm{x}_{j}-\bm{\mu}_l||^2}},
\end{equation}
where $\beta$ is the stiffness parameter that impacts the membership probability of each node. The best clustering solution is obtained by minimizing $J$, which differs from the conventional $k$-means since weighted squared errors are used in the cost function instead of squared errors  \cite{b8}. The result of the soft $k$-means algorithm will depend on the choice of $\beta$. We will discuss the choice of $\beta$ {when presenting} simulation results.

In order to minimize the objective function in (\ref{eq_object}), $z_{vj}$ must satisfy the following three constraints \cite{b8}.
\begin{enumerate}
\item Each node is assigned a membership probability between 0 and 1 for belonging to a cluster:
\begin{equation}
    z_{vj} \in [0,1], \ {v} = 1, \dots,k, \ j = 1,\dots,n.
\end{equation}
\item The sum of the membership probabilities for one node over all clusters is equal to 1:
\begin{equation}
    \sum_{v=1}^{k}z_{vj} = 1, \ j = 1, \dots, n.
\end{equation}
\item There will be at least one node with some non-zero membership probability for belonging to each cluster

\begin{equation}
    \sum_{j=1}^{n}z_{vj} > 0, \ v = 1,\dots,k.
\end{equation}
\end{enumerate}
By minimizing the objective function, we can calculate the cluster centers as \cite{b8}
\begin{equation}
\label{eq3}
\bm{\mu}_{v} = \frac{\sum_{j=1}^{n}z_{vj}\bm{x}_j}{\sum_{j=1}^{n}z_{vj}}.
\end{equation}

\begin{figure}[!t]
	\centering
	\includegraphics[width=3.5in]{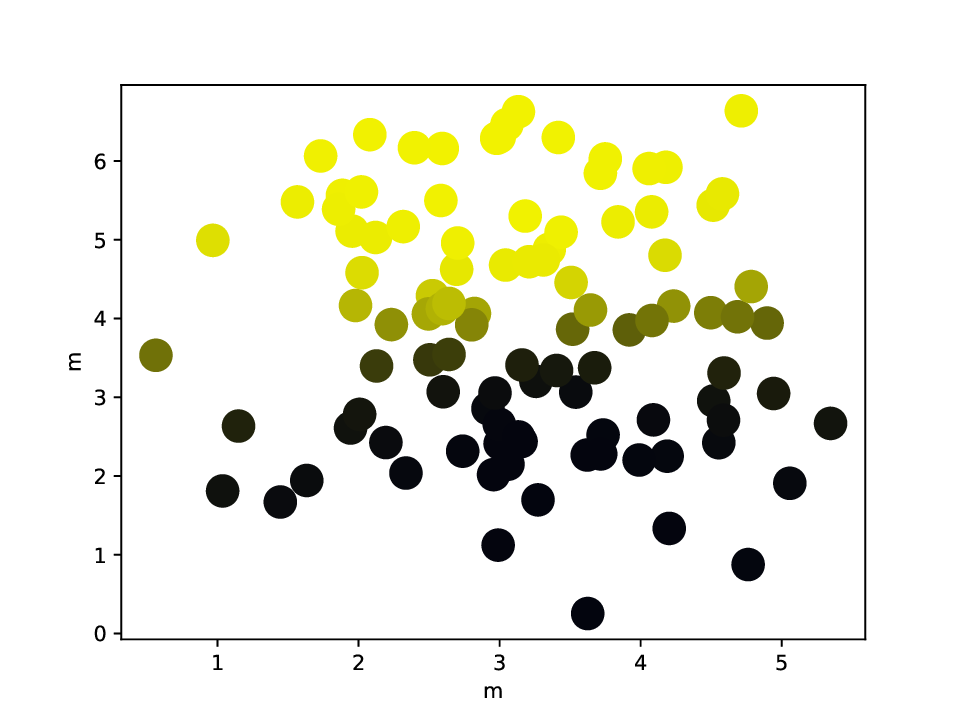}
	\caption{Example of soft $k$-means clustering.}
	\label{soft}
\end{figure}

The operations of the soft $k$-means algorithm can be summarized as follows: the algorithm calculates the membership probabilities and the cluster centers according to (2) and (6) in each round, respectively. If the changes of the membership probabilities $\bm{Z}$ or the cluster centers $\bm{M}$ are below given thresholds, the clustering process ends. Otherwise, the algorithm recalculates the new membership probabilities $\bm{Z}$ and the new cluster centers $\bm{M}$. If the algorithm does not converge after a given number of iterations, it will re-initiate by choosing new initial cluster centers. Fig. \ref{soft} shows an example of the clustering result of 100 nodes by the soft $k$-means algorithm.

\subsection{Kernel Density Estimation}

Non-parametric estimators are flexible for modeling probability density function (PDF) of data points. They have no fixed functional form and depend on data points to reach an estimate when compared to parametric estimators \cite{b26}. Non-parametric estimators can be classified into histogram-based and kernel-based estimation. A histogram-based estimator needs large data sets to guarantee convergence, and it cannot produce smooth continuous estimation curve \cite{b27}. KDE finds the distribution characteristics from data points without attaching any assumptions to data. It can ensure a smooth PDF approximation for given data points \cite{b29}. In KDE, the kernel function is centered at each data point, and it has the peak value at the data point location while decreasing in intensity with the distance from this location \cite{b29}.

\begin{figure*}[t!]
	\centering
	\subfigure[]{
		\begin{minipage}[h]{0.3\linewidth}
			\centering
			\includegraphics[width=2.1in]{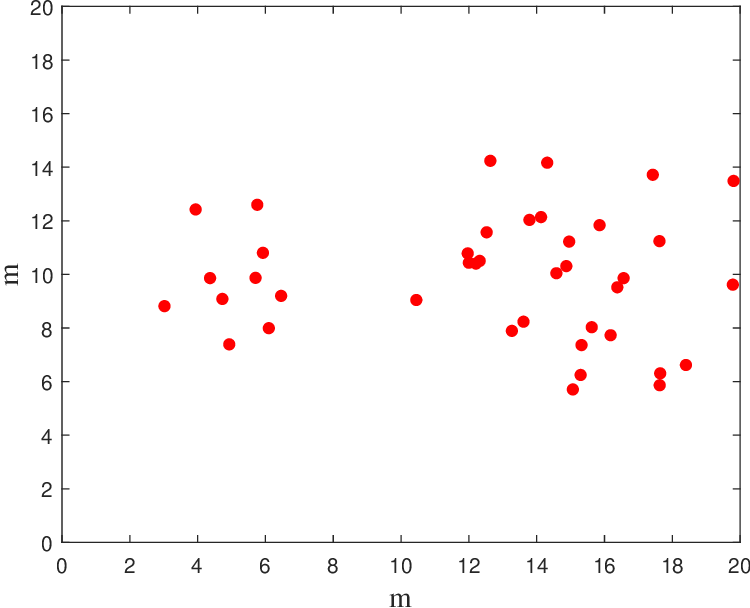}
			\label{kde}
		\end{minipage}
	}%
	\subfigure[]{
		\begin{minipage}[h]{0.3\linewidth}
			\centering
			\includegraphics[width=2.4in]{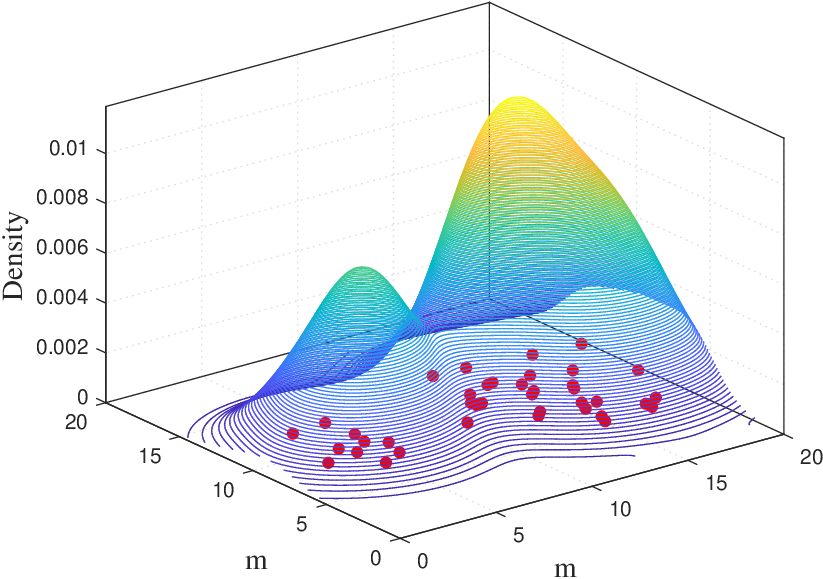}
			\label{kde2}
		\end{minipage}
	}%
	\subfigure[]{
		\begin{minipage}[h]{0.4\linewidth}
			\centering
			\includegraphics[width=2.3in]{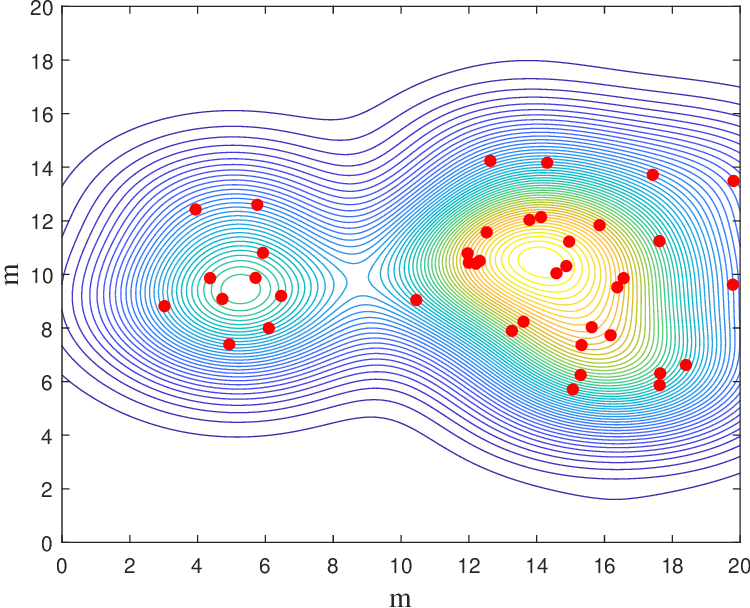}
			\label{kde3}
		\end{minipage}
	}
	\centering
	\caption{An example of KDE. (a) Nodes distribution. (b) 3-dimensional density contour of nodes in (a). (c) 2-dimensional density contour of nodes in (a).  }
	\label{fig22}
\end{figure*}

Using KDE, the PDF of the nodes' locations $\bm{X} = \{\bm{x}_1,\dots,\bm{x}_n\} \in \mathbb{R}^d$ is represented by a weighted sum of the kernel functions \cite{b30}
\begin{equation}
\label{kde100}
    \hat{f_h}(\bm{x}_i) = \frac{1}{nh^d}\sum_{t=1}^{n}\mathcal{K}\left(\frac{\bm{x}_t-\bm{x}_i}{h}\right),
\end{equation}
  where $h$ is the smoothing parameter called the bandwidth and it controls the size of the neighborhood around $\bm{x}_i, i\in{1,\dots,n}$. $\mathcal{K}(\cdot)$ is called the kernel function, which is defined in a $d$-dimensional space. The kernel function controls the weight given to $\bm{X}$ at each point $\bm{x}_i$ based on their proximity. To yield meaningful estimates, a kernel function should satisfy the following conditions \cite{b29}.
\begin{enumerate}
\item Normalization:
\begin{equation}
    \int_{\mathbb{R}^d}\mathcal{K}(\bm{u})d\bm{u} = 1.
\end{equation}
\item Symmetry:
\begin{equation}
    \mathcal{K}(\bm{-u}) = \mathcal{K}(\bm{u}).
\end{equation}
\item Non-negative and real-valued integrable:
\begin{equation}
    \mathcal{K}(\bm{u}) > 0.
\end{equation}
\end{enumerate}

A multivariate kernel function can be seen as a product of symmetric univariate kernel functions \cite{b32}
\begin{equation}
    \mathcal{K}(\bm{u}) = \prod_{j=1}^{d} \phi(u_{j}),
\end{equation}
where $u_{j}$ is the $j$th component of the $d$-dimensional vector $\bm{u}$, and $\phi(\cdot)$ is a univariate kernel function. In our proposed algorithm, we use the Gaussian kernel function due to its well-known properties \cite{b35}, which is defined as follows:
\begin{equation}
    \phi(u_{j}) = \frac{1}{\sqrt{2\pi}}\text{exp}\left(-\frac{u_{j}^2}{2}\right).
\end{equation}

Fig. \ref{fig22} is an example of KDE for a set of data. The set of discrete points is transformed into a smooth density map, as shown in Fig. \ref{fig22} (b), which displays its spatial distribution. The higher the PDF value in a location is, the higher the density is.

\subsection{``Clustering by Fast Search and Find of Density Peaks'' Algorithm}

CFSFDP is a new clustering algorithm proposed by Rodriguez and Laio \cite{b18}. It is based on the assumptions that cluster centers are surrounded by lower local density neighbors and they are at a relatively large distance from any nodes with a higher local density. This method needs to calculate two quantities for each node $i$: local density $\rho _i$ and distance $\delta_i$. The cluster centers are the nodes with higher local density and larger distance. For a set of nodes' locations $\bm{X} = \{\bm{x}_1,\bm{x}_2,\dots,\bm{x}_n \}$, and nodes' label set $\bm{I} = \{1,\dots,n\}$, the local density of a node $\bm{x}_i$ is defined as
\begin{equation}
    \rho_i = \sum_{i\neq j}\chi(d_{ij}-d_\text{c}),
\end{equation}
where
\begin{equation}
\chi(\alpha) = \left\{ \begin{array}{ll}
1, & \alpha < 0, \\
0, & \alpha \geq 0,
\end{array} \right.
\end{equation}
$d_{ij}$ is the distance between nodes $\bm{x}_i$ and $\bm{x}_j$, and $d_\text{c}$ is the cutoff distance. The choice of $d_\text{c}$ should yield an average number of neighbors around 1 to 2\% of the total number of nodes. In essence, $\rho_i$ can be seen as the number of nodes that are neighbor to node $\bm{x}_i$ in the range of $d_\text{c}$.

Two cases need to be considered in calculating a node's distance. If node $i$ has the highest density, then its distance $\delta_i$ is the maximum value of distances from node $i$ to all other nodes in $\bm{I}$. Otherwise, the distance of node $i$ is defined as the distance between node $i$ and its nearest neighbor having a higher density \cite{b36}. Specifically, the distance $\delta_i$ is expressed as
\begin{equation}
    \delta_i = \left\{ \begin{array}{ll}
        \max(d_{ij})_{j \in \bm{I}}, & \text{if }\rho_i \, \text{is maximum}, \\
        \min(d_{ij})_{j \in \bm{I}^{(i)}},  & \text{otherwise},\\
    \end{array} \right.
\end{equation}
\begin{equation}
    \bm{I}^{(i)} = \{t \in \bm{I}: \rho_t > \rho_i\},
\end{equation}
where $\bm{I}^{(i)}$ is the nodes' label set with node densities greater than $\rho_i$. After these two quantities are calculated, the cluster centers are selected from nodes with high values of both $\rho_{i}$ and $\delta_{i}$. Then, the CFSFDP algorithm assigns other remaining points to the nearest cluster center to form clusters. Specifically, if $\rho_i$ is large and $\delta_i$ is small for node $i$, it means node $i$ is close to the cluster center but not the center. On the other hand, if node $i$ has small $\rho_i$ and large $\delta_i$, it implies that the node is away from the cluster center \cite{b37}.

\begin{figure*}[!t]
  \centering
    \hspace{-0.1in}\subfigure[]{\includegraphics[width=0.35\textwidth]{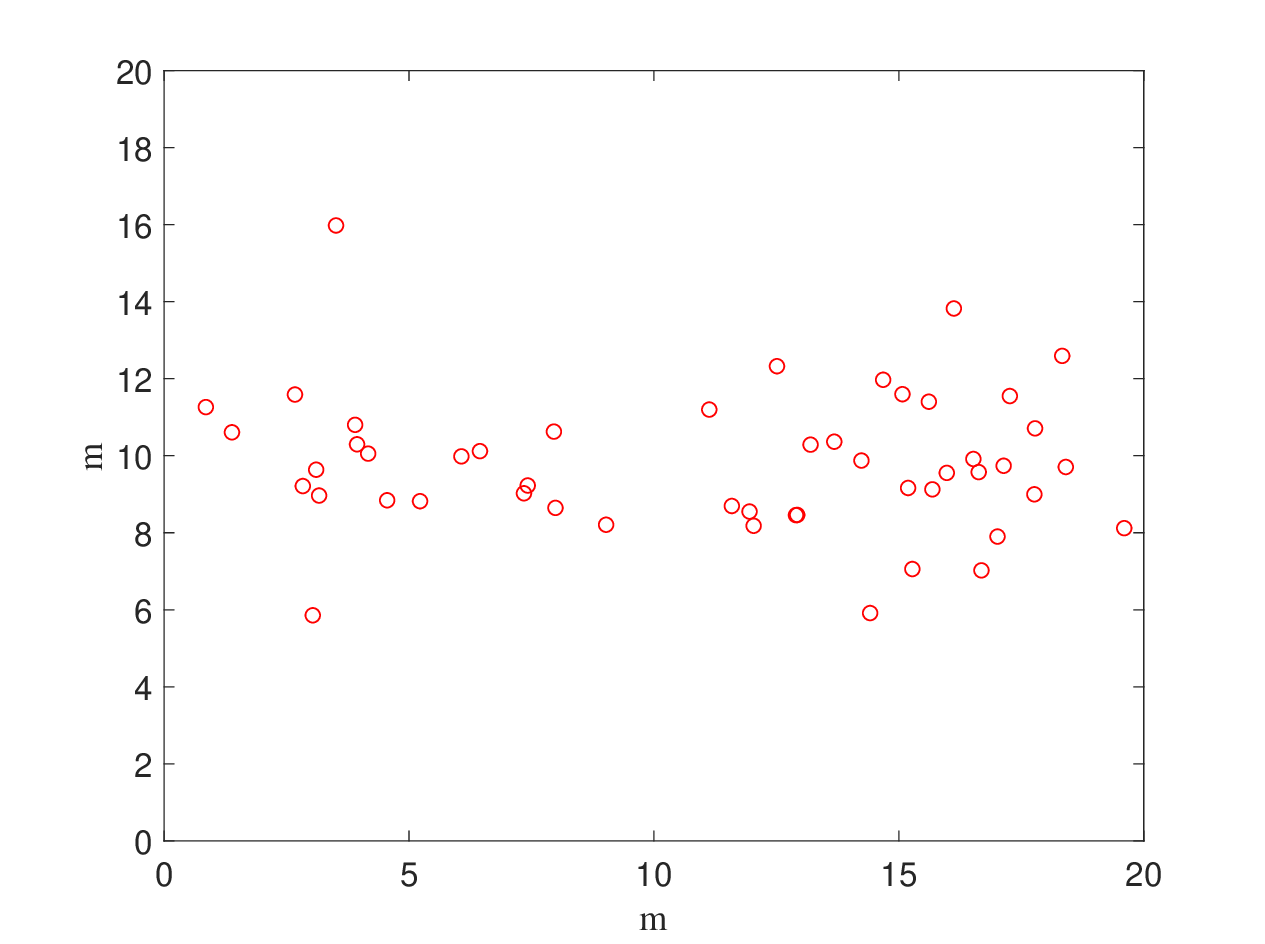}}\hspace{-0.2in}
	\subfigure[]{\includegraphics[width=0.35\textwidth]{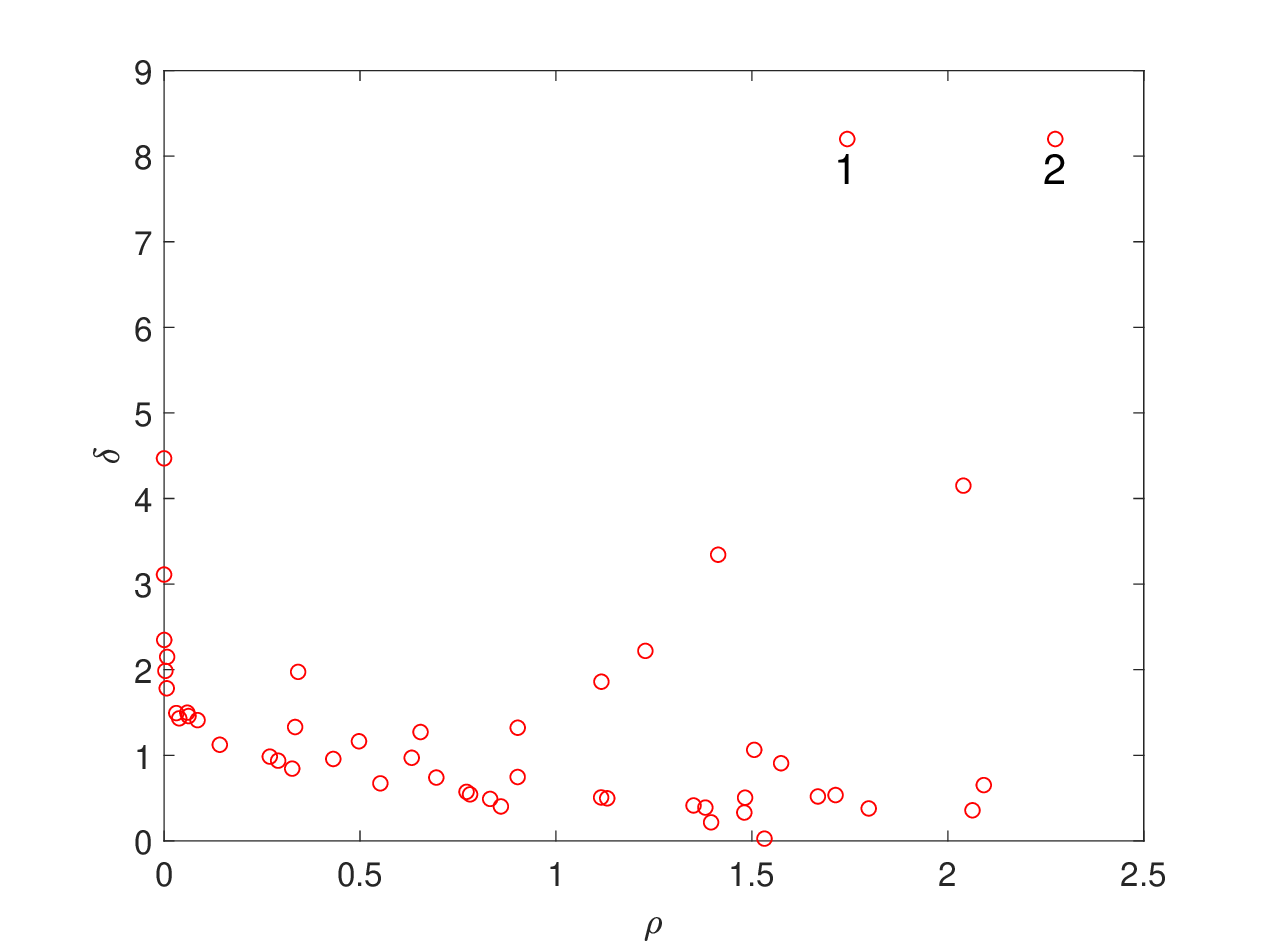}} \hspace{-0.2in}
	\subfigure[]{\includegraphics[width=0.35\textwidth]{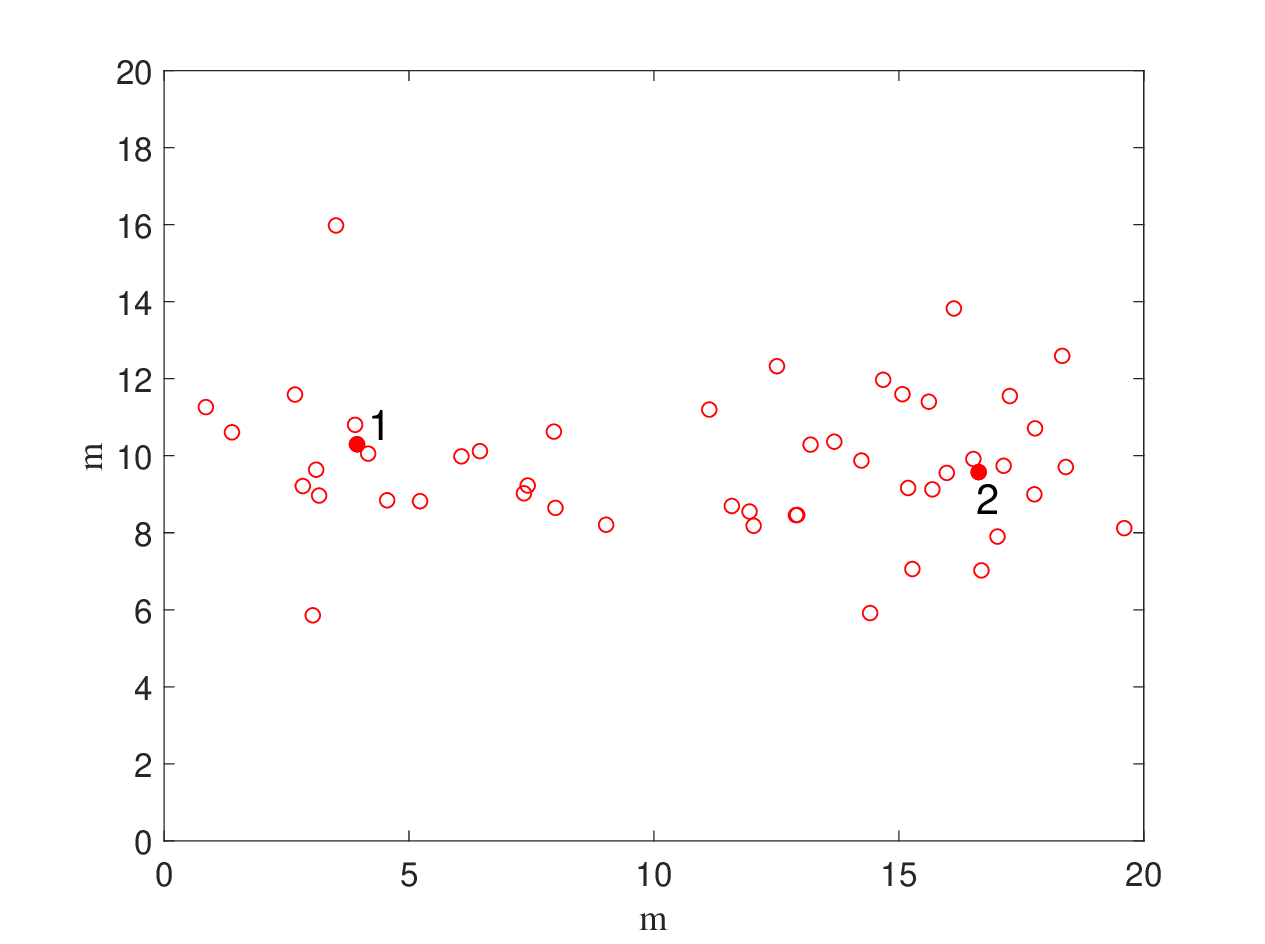}} \hspace{0in}
     \caption{CFSFDP in two dimensions. (a) Nodes distribution. (b) Decision graph for nodes in (a): X-coordinate is local density $\rho$, and Y-coordinate is $\delta$. (c) Two center nodes are determined.}
     \label{cfsfdp}
	\vspace{0in}
\end{figure*}

Fig. \ref{cfsfdp} (b) shows the plot of $\delta_i$ as a function of $\rho_i$ for each node in Fig. \ref{cfsfdp} (a). This representation is called the decision graph. According to the decision graph, we can have two nodes with higher values of both density $\rho$ and distance $\delta$. Hence, they can be chosen as cluster centers, as shown in Fig. \ref{cfsfdp} (c).

\section{Proposed IS-$\it{k}$-Means Algorithm}

\begin{figure}[!t]
	\centering
	\includegraphics[width=0.7\linewidth]{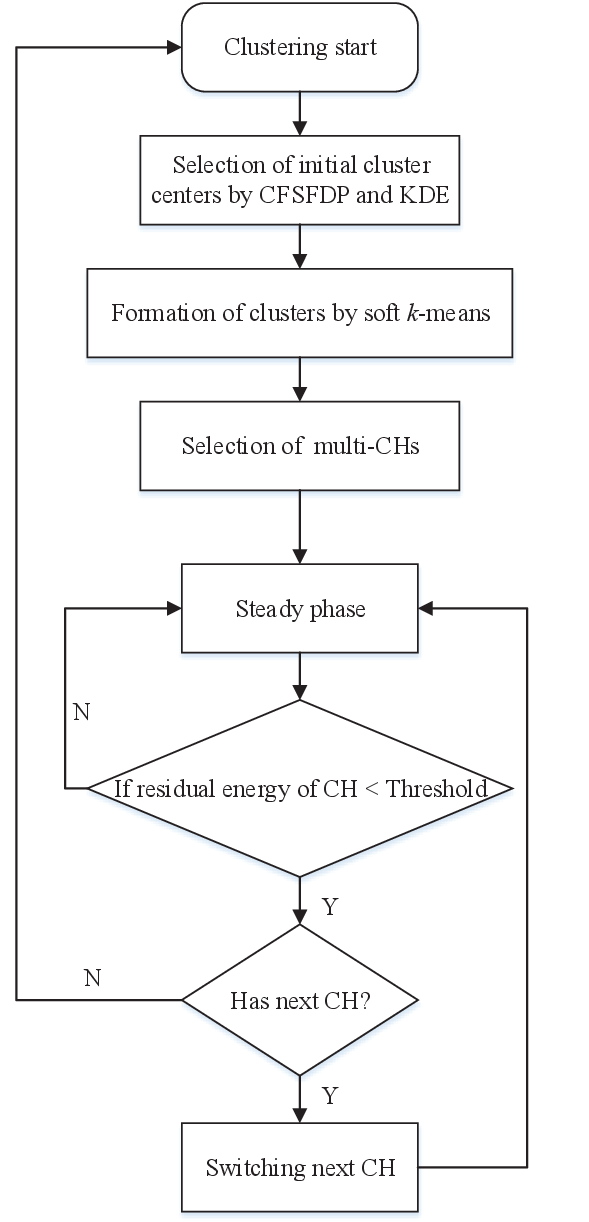}
	\caption{Flowchart of the proposed algorithm.}
	\label{flowchart}
\end{figure}

The proposed IS-$k$-means algorithm involves two phases: (i) set-up phase, and (ii) steady phases. During the set-up phase, each node broadcasts a HELLO message including its ID and location within the range of its coverage so that each node can acquire information of its neighbor nodes. Next, each node sends its information to the BS by the geographic multi-hop routing algorithm \cite{R. Zhang J. Pan} because it already knows the positions of its neighbor nodes. The BS runs the proposed IS-$k$-means algorithm according to the information received from all nodes. The proposed algorithm uses CFSFDP and KDE algorithms to optimize the selection of initial cluster centers of the soft $k$-means clustering method. Then, the soft $k$-means is used to form clusters and node reassigning scheme is employed to balance the numbers of nodes in different clusters. In order to balance the energy overhead of CHs, the multi-CHs scheme is utilized. After formulation of clusters and selection of CHs are completed, the BS broadcasts the results to all nodes by the restricted flooding method \cite{R. Zhang J. Pan}. Thus, each node can identify its role, e.g., CH or member node, and choose to join a corresponding CH if it is a member node. The steady phase is composed of many communication rounds. In each round $r$, member nodes collect and transmit data to CHs in their allotted time slots, and CHs aggregate the data and send it to the BS. When the energy of a CH is less then a threshold, it will broadcast a SWITCH message to activate the next candidate CH in the same cluster as the new CH and inform member nodes to send data to this new CH. If all CHs in a certain cluster are enabled sequentially, the last working CH will send a RESTART message to the BS to trigger re-clustering. The flowchart of the proposed IS-$k$-means algorithm is shown in Fig. \ref{flowchart}.

\subsection{Energy Model}

The first-order radio model  \cite{b10} is used to calculate the energy consumption of the network. The transmitter's energy consumption involves the transmitter circuitry and the power amplifier, while the energy consumption of the receiver accounts for the receiver circuitry. The free space and the multipath fading models are used in the transmitter power amplifier. If the distance between the transmitter and the receiver is less than a threshold, the power amplifier uses the free space model; otherwise, the multipath model is used \cite{b38}. The energy consumption of the transmitter and the receiver for transmitting an $l$-bit message can be calculated as follows \cite{b10}
\begin{equation}
E_{\text{T}}=\left\{ \begin{array}{ll}
lE_{\text{elec}} + l\varepsilon_\text{fs}d^2, & d \leq d_0, \\
lE_{\text{elec}} + l\varepsilon_\text{mp}d^4, & d > d_0 ,
\end{array} \right.
\end{equation}
\begin{equation}
E_\text{R} = lE_{\text{elec}},
\end{equation}
\begin{equation}
    d_0 = \sqrt{\frac{\varepsilon_\text{fs}}{\varepsilon_\text{mp}}},
\end{equation}
where $E_{\text{T}}$ is the dissipated energy in the transmitter and $E_\text{R}$ is the dissipated energy in the receiver. $E_{\text{elec}}$ is the dissipated energy per bit in both the transmitter circuitry and the receiver circuitry. $d$ is the transmission distance between the transmitter and the receiver. $d_0$ is the distance threshold. $\varepsilon_\text{fs}$ and $\varepsilon_\text{mp}$ represent the radio amplifier energy parameter of the free space and multipath fading models \cite{R. Zhang J. Pan}, respectively.

Because there are many rounds within the steady phase, the energy consumption of a CH in round $r$ can be calculated as
\begin{equation}
\label{22}
E_{\text{CH}}(r)=gcE_{\text{T}} + g(clE_{\text{DA}} + E_\text{R}),
\end{equation}
where $E_{\text{DA}}$ represents the dissipated energy of data aggregation and $c$ is the data aggregation ratio. The first term of the right hand side of (\ref{22}) is the energy consumption of a CH for sending aggregated data to the BS and the second term is the energy consumption of receiving and aggregating data of $g$ member nodes. The energy consumption of a member node sending data to its CH in round $r$ is
\begin{equation}
E_{\text{nonCH}}(r)=E_\text{T}.
\end{equation}
Hence, the residual energy of node $i$ in round $r$ can be computed by
\begin{eqnarray}
\label{24}
 E_i\left(r\right)=\left\{ \begin{array}{ll}
 E_i\left(r-1\right) - E_{\text{CH}}(r), & i \in \text{CHs}, \\
 E_i\left(r-1\right) - E_{\text{nonCH}}(r), & i \notin \text{CHs},
\end{array} \right.
\end{eqnarray}
where $E_i\left(r-1\right)$ is the residual energy of node $i$ in the $r-1$ round.

\subsection{Selection of Initial Cluster Centers}
 We use CFSFDP and KDE algorithms to determine the initial cluster centers as the input to the soft $k$-means clustering algorithm to produce a better clustering result. Because cluster centers are surrounded by neighbors with lower local density and they are at a relatively large distance from any points with a higher local density, they are selected by the maximum distance $\delta$ and relatively high local density $\rho$, which is illustrated in Fig. \ref{cfsfdp}. First, we calculate the density of each node and find the nodes' set $\bm{X}'$ with relatively high density $\bm{\rho}'$. Then, the distances $\delta$ among nodes in $\bm{X}'$ are computed. In order to choose cluster centers, we only choose nodes with relatively high density, and then we multiply their density $\rho_i$ and distance $\delta_i$ together as
 \begin{equation}
 \label{juli}
    \gamma_i = \rho_i \times \delta_i, \ i \in \{1,\dots,m\},
 \end{equation}
 where  $m$ is the number of nodes with relatively high density. Since each initial cluster center node should have a high $\gamma$ value, we choose nodes with relatively large $\gamma$ value as the initial cluster centers. In addition, the value of $k$ is equal to the number of the initial cluster centers. Algorithm \ref{alg1} describes the detailed steps.

\begin{figure}[t!]
	\centering
	\includegraphics[width=0.7\linewidth]{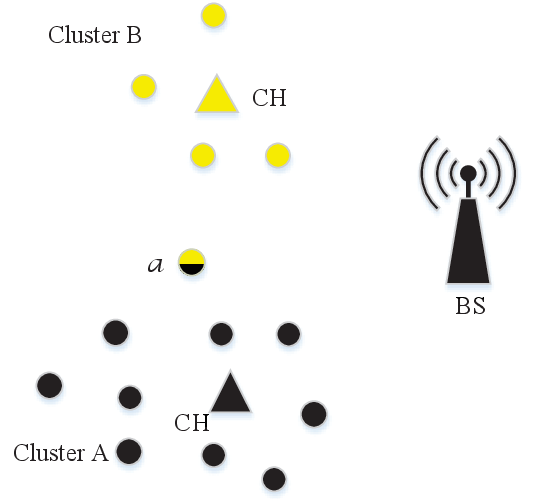}
	\caption{A node at the boundary of two clusters.}
	\label{fig5}
\end{figure}

\begin{algorithm}[t!]
	\caption{Selection of initial cluster centers}
	\label{alg1}
	\begin{algorithmic}[1]
		\renewcommand{\algorithmicrequire}{\textbf{Input:}}
		\renewcommand{\algorithmicensure}{\textbf{Output:}}
		\REQUIRE $\bm{X} = \{\bm{x}_1,\dots,\bm{x}_n\}$\\
	    \ENSURE  Initial cluster centers: $\bm{M}$\\
		\FOR {i = $\mathrm{1}: n$}
		\STATE calculate $\rho_i$
		\ENDFOR
		\STATE $\bm{\rho} = \{\rho_1,\dots,\rho_n\}$
		\STATE choose nodes with local maximum density $\bm{X}' = \{\bm{x}_1,\dots,\bm{x}_m\}$ and get their density set $\bm{\rho}' = \{\rho_1,\dots,\rho_m\}, m < n$
		\FOR{i = $\mathrm{1}:m$}
		\STATE calculate $\delta_i$
		\ENDFOR
		\STATE $\bm{\Delta} = \{\delta_1,\dots,\delta_m\}$
		\STATE calculate $\gamma_i$ by (\ref{juli}) to determine the initial cluster centers
		\RETURN $\bm{M} = \{\bm{\mu}_1,\dots,\bm{\mu}_k\}$\\
	\end{algorithmic}
\end{algorithm}

\subsection{Cluster Formation}
\begin{algorithm}[t!]
	\caption{Cluster formation}
	\label{alg2}
	\begin{algorithmic}[1]
		\renewcommand{\algorithmicrequire}{\textbf{Input:}}
		\renewcommand{\algorithmicensure}{\textbf{Output:}}
		\REQUIRE $\bm{M} =
		\{\bm{\mu}_1,\dots,\bm{\mu}_k\}$, $ \bm{X} = \{\bm{x}_1,\dots,\bm{x}_n\}$, the maximum number of iterations $r_\text{max}$
		\ENSURE  $k$ clusters \\
		\FOR{$r$ = $\mathnormal{1} : r_\text{max}$}
		    \FOR{{$v$ = $\mathnormal{1}:k$}}
		        \FOR{{$j$ = $\mathnormal{1}:n$}}
		           \STATE{{$z'=0$}}
                    \FOR{{$l$ = $\mathnormal{1}:k$}}
				       \STATE
				       {${z' = z' + e^{-\beta||\bm{x}_j-\bm{\mu}_l||^2}}$}
				   \ENDFOR\\
				       \STATE {$z_{vj} = \frac{e^{-\beta||\bm{x}_j-\bm{\mu}_v||^2}}{z'}$}
			    \ENDFOR
		    \ENDFOR
			\STATE ${\bm{Z}_r}={
				\left[ \begin{array}{cccc}
				z_{11} & z_{12} & \cdots & z_{1n}\\
				\vdots & \vdots & \vdots & \vdots\\
				z_{k1} & \cdots & \cdots & z_{kn}
				\end{array}
				\right ]}$
			\FOR{$v = \mathnormal{1}:k$}
				\STATE {$\bm{\mu}_v = \frac{\sum_{j=1}^{n}z_{vj}\bm{x}_{j}}{\sum_{j=1}^{n}z_{vj}}$}
			\ENDFOR
		\ENDFOR
		\STATE final membership probabilities ${\bm{Z}_{r_{\text{max}}}}$
		\FOR{j = $\mathnormal{1}:n$}
		\STATE assign node $j$ to cluster with the highest probability according to $\bm{Z}_{r_{\text{max}}}$
		\ENDFOR
		\STATE $k$ clusters $\bm{C} = \{\bm{c}_1,\dots,\bm{c}_k\}$
		\FOR{j = $\mathnormal{1}:n$}
			\STATE reassign node $j$ located on the border to different cluster
		\ENDFOR
		\RETURN $k$ new clusters $\bm{C}' = \{\bm{c}'_{1},\dots,\bm{c}'_{k}\}$
	\end{algorithmic}
\end{algorithm}

Some $k$-means-based algorithms form clusters according to the distances between normal nodes and CHs, such as distributed $k$-means clustering algorithm \cite{b19} and improved $k$-means cluster-based routing \cite{b20}. These $k$-means-based algorithms can easily lead to a large gap in the number of nodes in different clusters in WSNs and may cause unbalanced energy consumption of CHs. Hence, compared with these $k$-means-based clustering algorithms, our proposed IS-$k$-means algorithm uses the soft $k$-means clustering algorithm to address this problem. Each node can be a member of more than one clusters at the same time according to membership probabilities in the soft $k$-means. However, member nodes need to join only one cluster with the highest membership probability at a time. Some boundary nodes may have similar probabilities to join multiple clusters. After the convergence of our proposed IS-$k$-means algorithm, we may reassign nodes to different clusters to balance the number of nodes per cluster. For example, node $a$ is at the edge of two clusters and it has a higher probability to join cluster A, as illustrated in Fig. \ref{fig5}. Before reassigning node $a$, cluster A already has 10 member nodes and cluster B has 5 member nodes. Since all member nodes send messages to their CH, CH of cluster A will deal with more information from its member nodes. In order to balance the energy consumption of CHs, it is better to reassign node $a$ to cluster B. Reassigning node $a$ from cluster A to cluster B may increase slightly the energy consumption of transmitting messages between node $a$ and its CH because the transmission distance $d$ is increased. However, this slight increase of the transmission energy consumption is negligible as compared to the total energy consumption in CH. If the difference of the probabilities of a node belonging to two clusters is less than a certain threshold, it will join the cluster with low density. If a node is at the boundary of three or more clusters, the proposed algorithm only choose the first two maximum probabilities and follow the same rule. Algorithm \ref{alg2} outlines the cluster formation algorithm.

\subsection{Selection of Multi-CHs}

Normally, the numbers of nodes in different clusters are different in WSNs. If only one CH is selected in each cluster, CH will consume too much energy to deal with the information from its member nodes in a high density cluster, which will cause its death too early. Hence, our proposed IS-$k$-means algorithm designs a scheme of multi-CHs. The number of CHs is not fixed in each cluster, and it is determined by the number of nodes per cluster. The larger the number of nodes in a cluster is, the higher the number of CHs will be. The remaining energy of nodes and distances between nodes and their cluster centers are considered in choosing CHs. Nodes close to their cluster center and having higher residual energy than the average energy of the cluster can become CHs. We define a matrix $\textbf{CHs} =\{ {\textbf{CH}_1,\dots, \textbf{CH}_k}\}$, which is composed of all CHs of $k$ clusters, and $\textbf{CH}_v, 1 \leq v \leq k$, represents the set of CHs of cluster $v$. The total remaining energy of cluster $v \in \{1,\dots,k\}$ can be computed as
\begin{equation}
E_v = \sum_{i=1}^{S_v}E_i\left(r\right),
\end{equation}
where $S_v$ is the size of cluster $v$, $E_i\left(r\right)$ is the residual energy of node $i$ in current round $r$, which can be obtained from (\ref{24}).

\begin{figure}[t!]
    \centering
    \includegraphics[scale=0.7]{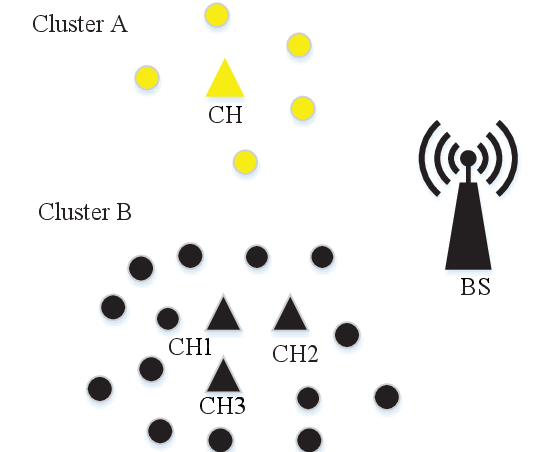}
    \caption{Multi-CHs scheme.}
    \label{fig7}
\end{figure}

\begin{algorithm}[t!]
	\caption{Selection of multi-CHs}
	\label{alg3}
	\begin{algorithmic}[1]
		\renewcommand{\algorithmicrequire}{\textbf{Input:}}
		\renewcommand{\algorithmicensure}{\textbf{Output:}}
		\REQUIRE  $\bm{C}' = \{\bm{c}'_{1},\dots,\bm{c}'_{k}\}$
		\ENSURE \textbf{CHs}
		\FOR{$v = \mathnormal{1}:k$}
			\STATE calculate the size of cluster $\bm{c}'_{v}$: $S_v$
			\STATE calculate average energy of cluster $E_{\text{ave}_v}$
			\STATE $p = \frac{S_v}{\text{constant}}$, the number of CHs of cluster $v$
			\FOR{i = $\mathnormal{1}:S_v$}
				\STATE Iterate $\bm{x}_i$ from near the center of cluster
				\IF{$E_i > E_{\text{ave}_v}$ and $p > 0$}
					\STATE $p = p - 1$
					\STATE $\textbf{CH}_v\left(p\right) = \bm{x}_i$
				\ENDIF
			\ENDFOR
		\ENDFOR
		\RETURN $\textbf{CHs} = \{\textbf{CH}_1,\dots,\textbf{CH}_k \}$, CHs of $k$ clusters
	\end{algorithmic}
\end{algorithm}

The average energy of cluster $v$ is calculated as
\begin{equation}
    E_{\text{ave}_v} = \frac{E_v}{S_v}.
\end{equation}

As an example, since the number of nodes in cluster B is around 3 times that of cluster A in Fig. \ref{fig7}, cluster B will have three CHs if only one CH is selected in cluster A. After the number of CHs is determined, the nodes which have larger remaining energy and close to the cluster center are selected as CHs. This multi-CHs scheme can balance the energy consumption of CHs per cluster in WSNs, and is summarized in Algorithm \ref{alg3}.

After the set of CHs and clusters are determined, the first node in each $\textbf{CH}_v$ is selected as the current CH in that cluster and the BS notifies all member nodes to join the cluster to which they belong. CHs broadcast time division multiple access schedules to their member nodes for transmitting data in different time slots to avoid data collision. Then, the network enters the steady phase and begins to exchange data between normal nodes and their CHs.

 \begin{algorithm}[!t]
	\caption{Switching to a next CH}
	\label{alg4}
	\begin{algorithmic}[1]
		\renewcommand{\algorithmicrequire}{\textbf{Input:}}
		\renewcommand{\algorithmicensure}{\textbf{Output:}}
		\REQUIRE CHs of $k$ clusters, $\textbf{CHs} = \{\textbf{CH}_1,\dots,\textbf{CH}_k \}$
		\ENSURE Next CH
		\STATE 	Current round
		\FOR{$v = \mathnormal{1}:k$}
			\STATE $T = \frac{\text{residual energy of}\,  \textbf{CH}_v(p) \,\text{{in current round}}}{\text{residual energy of}\,  \textbf{CH}_v(p) \,\text{{in last round}}}$
			\IF{$T < \text{Threshold}$}
				\IF{$\textbf{CH}_v$ has $\textbf{CH}_v(p+1)$}
					\STATE switch to $\textbf{CH}_v(p+1)$
				\ELSE
					\STATE re-clustering
				\ENDIF
			\ENDIF
		\ENDFOR
	\end{algorithmic}
 \end{algorithm}

\begin{figure*}[!t]
  \centering
    \hspace{-0.2in}\subfigure[]{\includegraphics[width=0.35\textwidth]{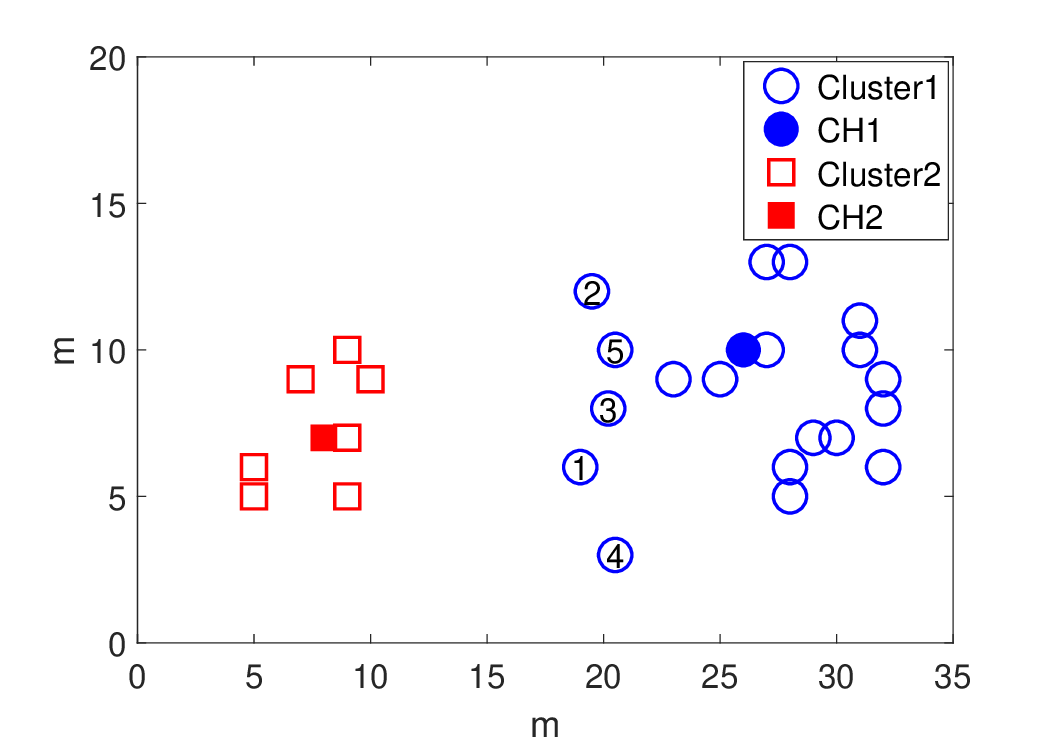}}\hspace{-0.2in}
	\subfigure[]{\includegraphics[width=0.35\textwidth]{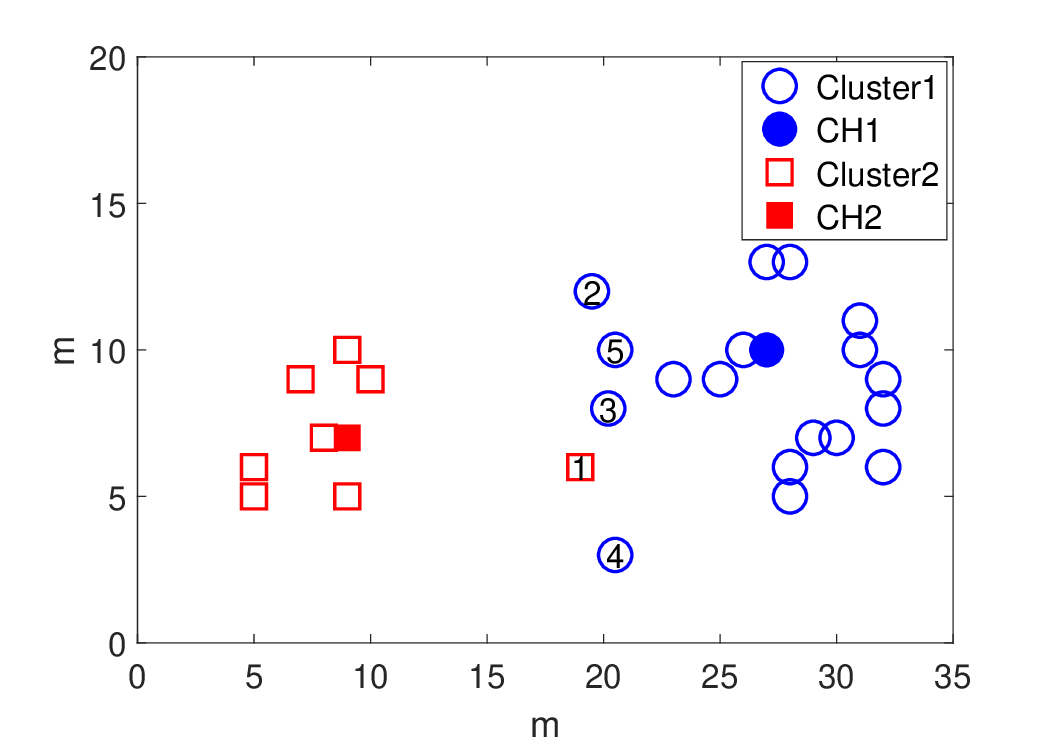}} \hspace{-0.2in}
    \subfigure[]{\includegraphics[width=0.35\textwidth]{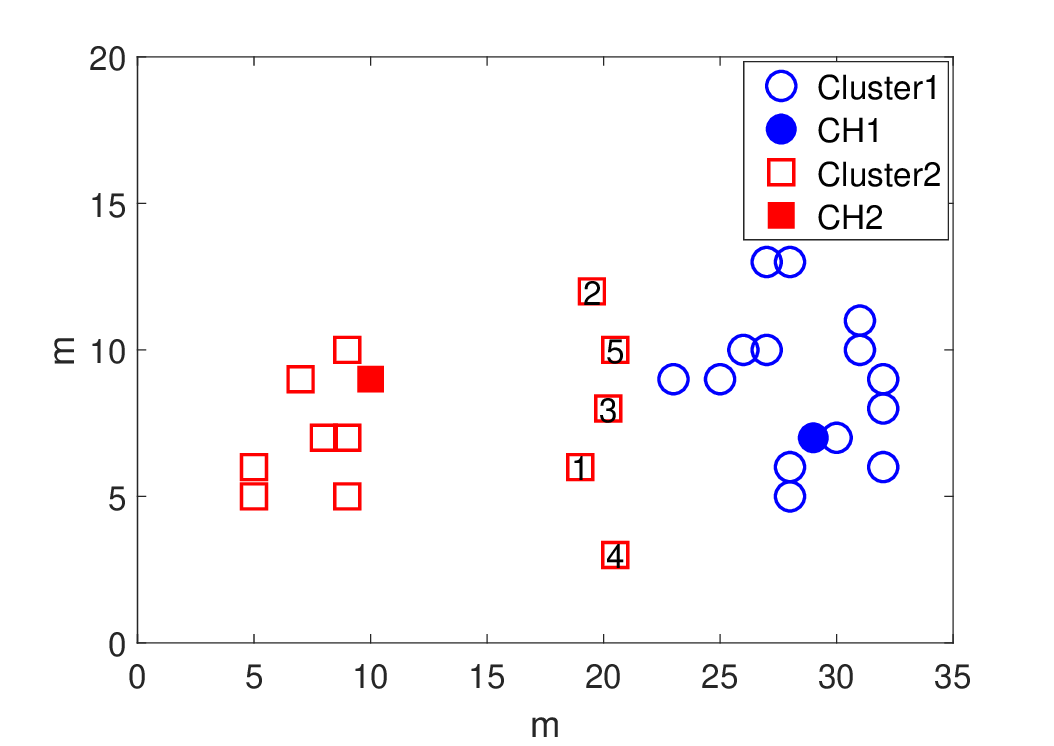}}\hspace{0in}
     \caption{Comparison of different clustering results, $\beta = 0.2$. (a) \textit{k}-means clustering result. (b) Soft \textit{k}-means clustering result. (c) IS-\textit{k}-means clustering result.}
     \label{reassign}
	\vspace{0in}
\end{figure*}

\begin{figure*}[!t]
  \centering
    \hspace{-0.1in}\subfigure[]{\includegraphics[width=0.4\textwidth]{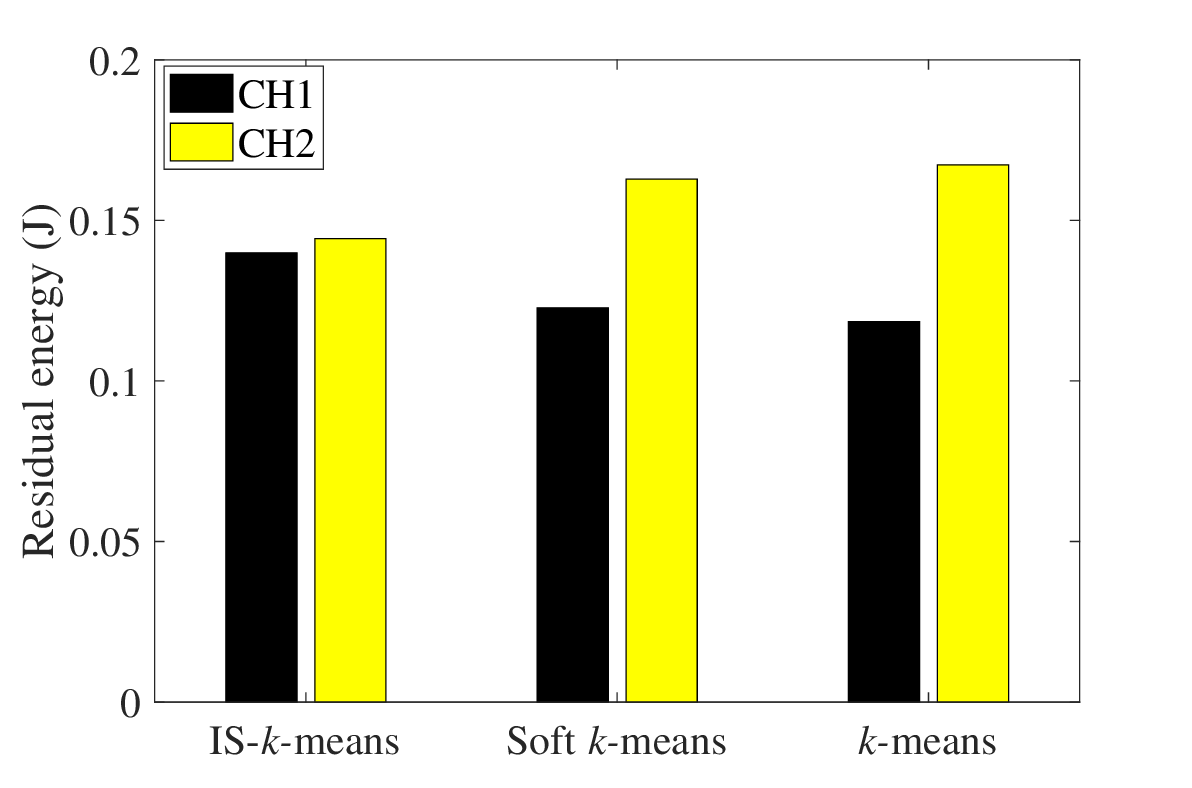}\label{Residual energy of CHs after 5 rounds}}\hspace{0in}
	 \subfigure[]{\includegraphics[width=0.4\textwidth]{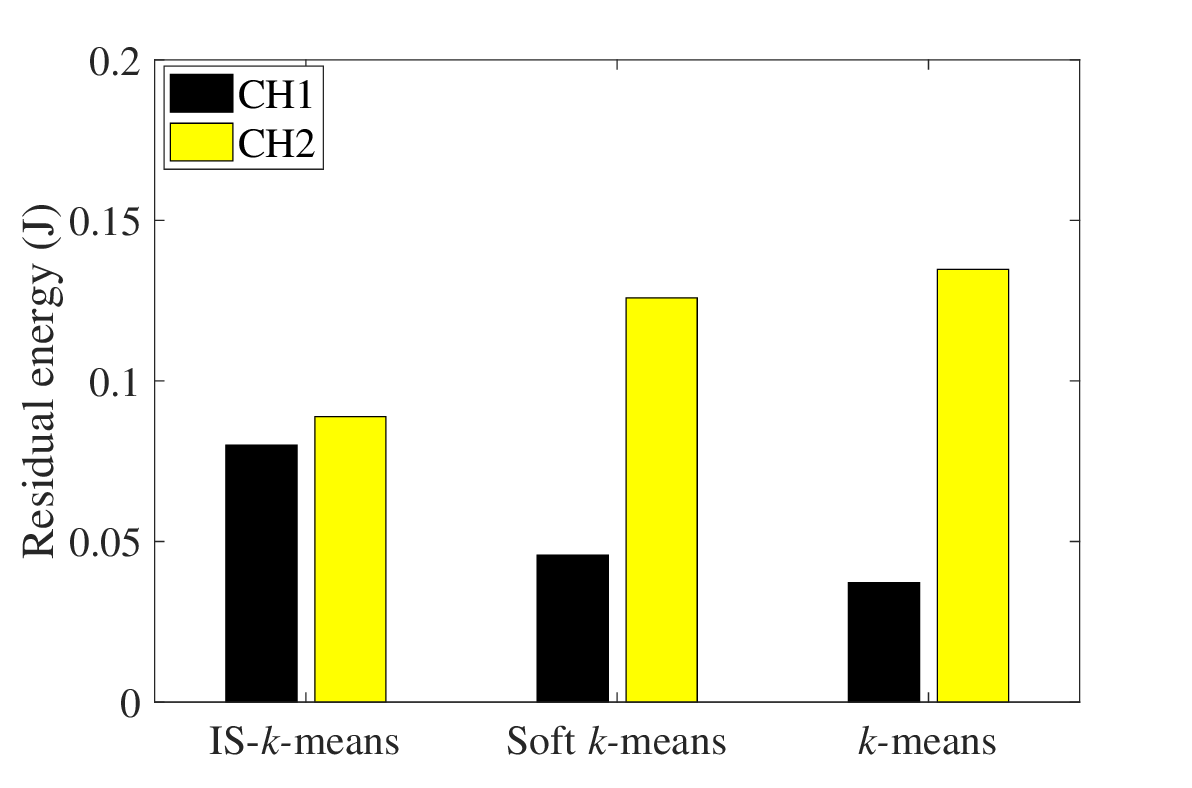}\label{Residual energy of CHs after 10 rounds}} \hspace{0in}
     \caption{Comparison of residual energy of CHs. (a) Residual energy of CHs after 5 rounds. (b) Residual energy of CHs after 10 rounds.}
     \label{reassignenergy}
	\vspace{0in}
\end{figure*}

\subsection{Switching to a Next CH}
For balancing the energy consumption of CHs, if the energy consumption ratio of a current CH of any cluster is below a threshold value, the next candidate CH in that cluster is enabled. Until all CHs in a given cluster are executed, the algorithm starts re-clustering. The specific steps are described in Algorithm \ref{alg4}.

\subsection{Complexity Analysis}
The run-time complexity of the proposed IS-$k$-means algorithm mainly involves three phases. In the phase of selecting initial cluster centers, IS-$k$-means needs $o(n^2)$ operations \cite{W. Zhang and J. Li} to execute CFSFDP and KDE to calculate the nodes' densities and distances where $n$ is the number of nodes. Then, the algorithm requires $o(nk^2r_{\text{max}})$ operations \cite{T. C. Havens} to execute the soft $k$-means and $o(2n)$ operations to assign nodes to form final clusters. Because the selection of initial cluster centers has been optimized, the algorithm converges quickly and the value of $r_{\text{max}}$ is very small. In the third phase, the algorithm needs $o(n)$ operations to select CHs. Thus, the overall time complexity  of the proposed IS-$k$-means algorithm is $o(n^2 + nk^2r_{\text{max}} + 3n)$ operations. Obviously, the time complexity of the IS-$k$-means depends mainly on the execution time of the first phase. The time complexity of the soft $k$-means algorithm is $o(nk^2r_{\text{max}})$ operations \cite{T. C. Havens}, which is lower than that of our proposed IS-$k$-means algorithm. However, the higher complexity of the proposed IS-$k$-means algorithm can be well justified by its ability to better balance the energy consumption of nodes. As for the memory requirement, the proposed algorithm needs $o(n)$ memory units to store nodes first. Then, it costs $o(n)$ memory units \cite{M. Wang} to store $\rho$ and $\delta$ in the phase of selecting initial cluster centers. Then, $o(nk)$ memory units are required to store membership probabilities in the phase of cluster formation. Hence, the total storage requirement of the proposed algorithm is $o(2n + nk)$ memory units.

\section{Experiment Results and Analysis}

\subsection{Simulation Settings}
To evaluate the performance of the proposed algorithm, we consider two different scenarios. In Scenario 1, the network size is $100 \, \text{m} \times  100\, \text{m}$ and the BS is located at $(50\, \text{m},150\, \text{m})$. Scenario 2 has the size of $200 \, \text{m} \times 200\, \text{m}$ with the BS at location $(100\, \text{m},200\, \text{m})$. The main simulation parameters are selected as in \cite{b2} and listed in Table \ref{table1}. The experiments are implemented using MATLAB R2017b.

\begin{table}[t!]
	\centering
	\caption{Simulation parameters}
	
	\label{table1}
	\begin{tabular}{p{4cm}p{4cm}}
		\Xhline{1pt}
		\textbf{Parameter} & \textbf{Value}\\
		\hline
		Area & $100\,\text{m} \times 100\,\text{m}$, {$200\,\text{m} \times 200\,\text{m}$}  \\
		\hline
		BS coordinates & (50\,\text{m}, 150\, \text{m}), {(100\,\text{m}, 200\, \text{m})}\\
		\hline
		Initial energy & 0.2\,J, 1\,J\\
		\hline
		Packet length & 4000\,bits\\
		\hline
		Control length & 100\,bits\\
		\hline
		$E_\text{T}$ & 50\,nJ/bit \\
		\hline
		$E_\text{R}$ & 50\,nJ/bit \\
		\hline
		$\varepsilon_\text{fs}$ & 10\,pJ/bit/$\text{m}^2$\\
		\hline
		$\varepsilon_\text{mp}$ & 0.0013\,pJ/bit/$\text{m}^4$\\
		\hline
		$E_\text{DA}$ & 5\,nJ/bit\\
		\hline
		$d_0$ & 88\,m\\
		\hline
		Number of sensor nodes & 28, 100\\
		\hline
		{Maximum communication range} & {$250\, \text{m}$\cite{C. Lipi}}\\
		\Xhline{1pt}
	\end{tabular}
\end{table}

\subsection{Nodes Reassigning of Improved Soft $k$-Means Analysis}

In this subsection, we will show the advantage of the node reassigning scheme incorporated in the proposed IS-$k$-means algorithm to balance the energy consumption of CHs. A total of 28 sensor nodes are randomly distributed in scenario 1. First, we use the $k$-means clustering method to classify these nodes and obtain two clusters, as shown in Fig. \ref{reassign} (a). It is found that cluster 1 contains 20 nodes, which is quite larger than the number of nodes in cluster 2. As a result, CH of cluster 1 will be exhausted much earlier than that of cluster 2. Fig. \ref{reassign} (b) shows the clustering result of the soft $k$-means algorithm. In Section \uppercase\expandafter{\romannumeral3}, we define $\beta$ as the stiffness parameter, which represents the tightness of a node belonging to a cluster. Setting $\beta = 0.2$, we can find that the nodes at the edge of two clusters having similar membership probabilities belonging to these two clusters, such as node 1, node 2, node 3, node 4, and node 5, as shown in Table \ref{table2}. Furthermore, if the value of $\beta$ changes, the probabilities also will change.  When $\beta = 1$, all five nodes belong to the clusters with higher probabilities when compared to the case where $\beta = 0.2$. In our proposed algorithm, we set $\beta = 0.2$ in the following simulations. According to the rule of node reassigning, node 2, node 3, node 4, and node 5 are reassigned to cluster 2 from cluster 1 as shown in Fig. \ref{reassign} (c), which balances the energy overhead of CHs in these two clusters. The residual energy of CHs, computed by (\ref{24}), in each round could be used to check the advantage of this scheme. Fig. \ref{reassignenergy} (a) and Fig. \ref{reassignenergy} (b) compare the residual energy of CHs among $k$-means, soft $k$-means, and IS-$k$-means after 5 rounds and 10 rounds, respectively. The IS-$k$-means algorithm achieves an equilibrium of energy consumption in both CHs when compared to the $k$-means and the soft $k$-means algorithms.

\begin{table}[t!]
	\centering
	\caption{Probabilities comparison}
	\label{table2}
	 \begin{tabular}{p{0.8cm}<{\centering}|p{1.04cm}<{\centering}|c|c|c|c|c}
		\Xhline{1pt}
		\multicolumn{2}{c|}{Probability} & Node 1 & Node 2 & Node 3 & Node 4 & Node 5\\
		\hline
		\multirow{2}*{$\beta = 0.2$} & Cluster 1 & 0.4852 & 0.5537 & 0.5684 & 0.6125 & 0.6120 \\
		\cline{2-7}
		& Cluster 2 & 0.5148 & 0.4463 & 0.4316 & 0.3875 & 0.3880
		\\
		\hline
		\multirow{2}*{$\beta = 1$} & Cluster 1 & 0.0438 & 0.9787 & 0.9860 & 0.9919 & 0.9992 \\
		\cline{2-7}
		& Cluster 2 & 0.9562 & 0.0213 & 0.014 & 0.0081 & 0.0008\\
		\Xhline{1pt}
	\end{tabular}
\end{table}

\subsection{Network Lifetime}

To test the performance of the proposed IS-$k$-means algorithm, we compare it with KM-LEACH  \cite{b5}, VLEACH  \cite{b6}, LEACH  \cite{b10}, $k$-means \cite{b4}, EECPK-means  \cite{b15}, and EB-CRP \cite{R. Yarinezhad and S. N. Hashemi} with the same parameters shown in Table \ref{table1}. Here, we state two things about the implementation of the EB-CRP algorithm in our experiment. First, the original EB-CRP algorithm does not need to select the CHs because the authors consider a certain number of gateways with enough energy to act as CHs in WSN. However, our implemented EB-CRP algorithm needs to select CHs randomly from all sensor nodes because the network considered in our simulation contains only sensor nodes with the same initial energy and functionality. In order to have a fair comparison, we set the number of CHs in EB-CRP to be the same as that in our proposed algorithm. Thus, the location of CHs may be different in each steady-state phase because all nodes have the same chance to be CHs. Second, the steady-state phase of the original EB-CRP algorithm is composed of pre-specified 75 rounds. This is quite reasonable because the authors set the initial energy of CHs to be 10 J, which can maintain a high number of communication rounds. However, considering the limited energy of CHs in our simulation, each steady-state phase is composed of 20 rounds in our implemented EB-CRP algorithm, which can achieve the best results for the EB-CRP algorithm.

\begin{figure*}[!t]
	\centering
	 \hspace{-0.3in}\subfigure[]{\includegraphics[width=0.54\textwidth]{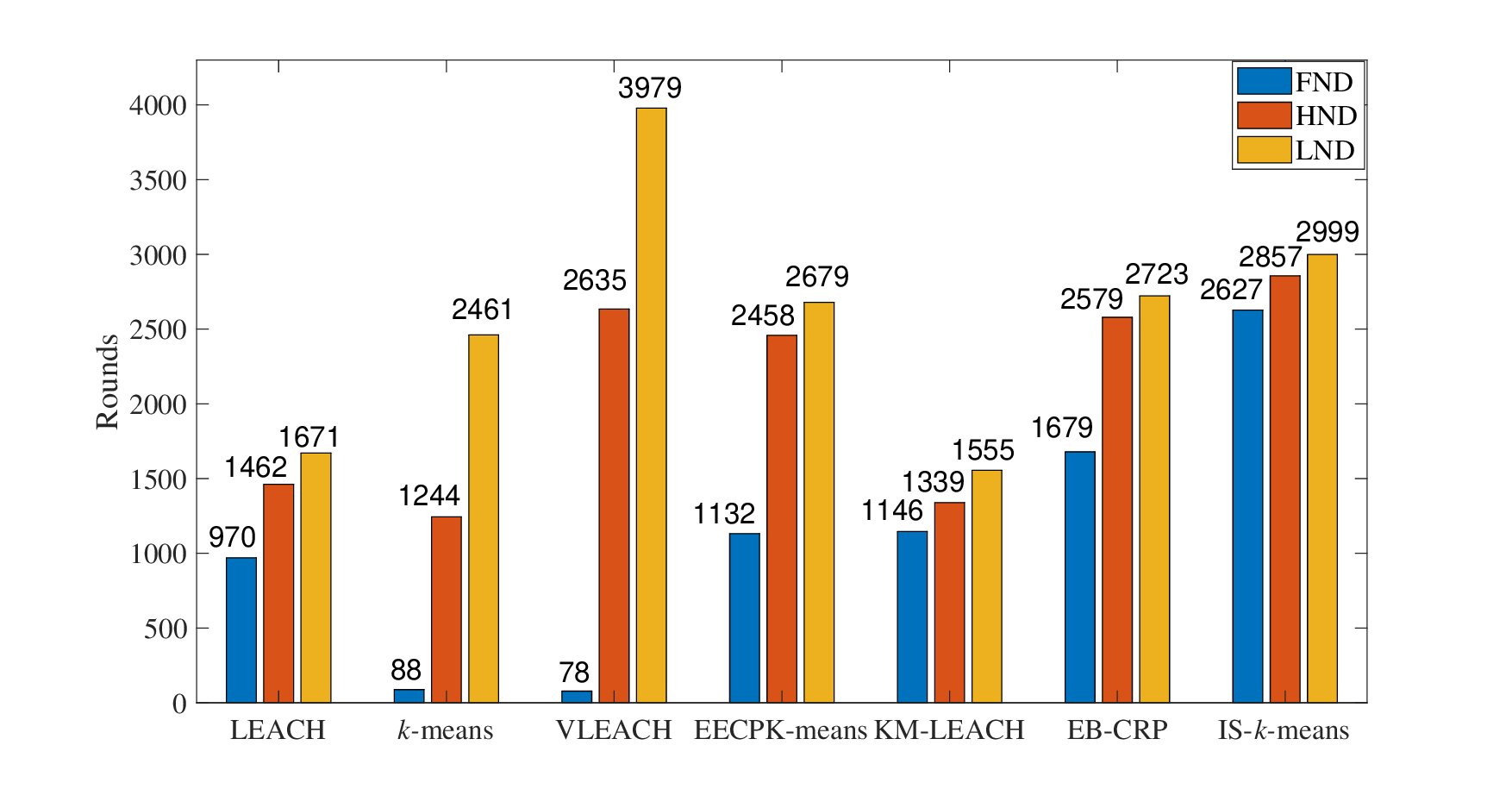}}
	 \hspace{-0.4in}\subfigure[]{\includegraphics[width=0.54\textwidth]{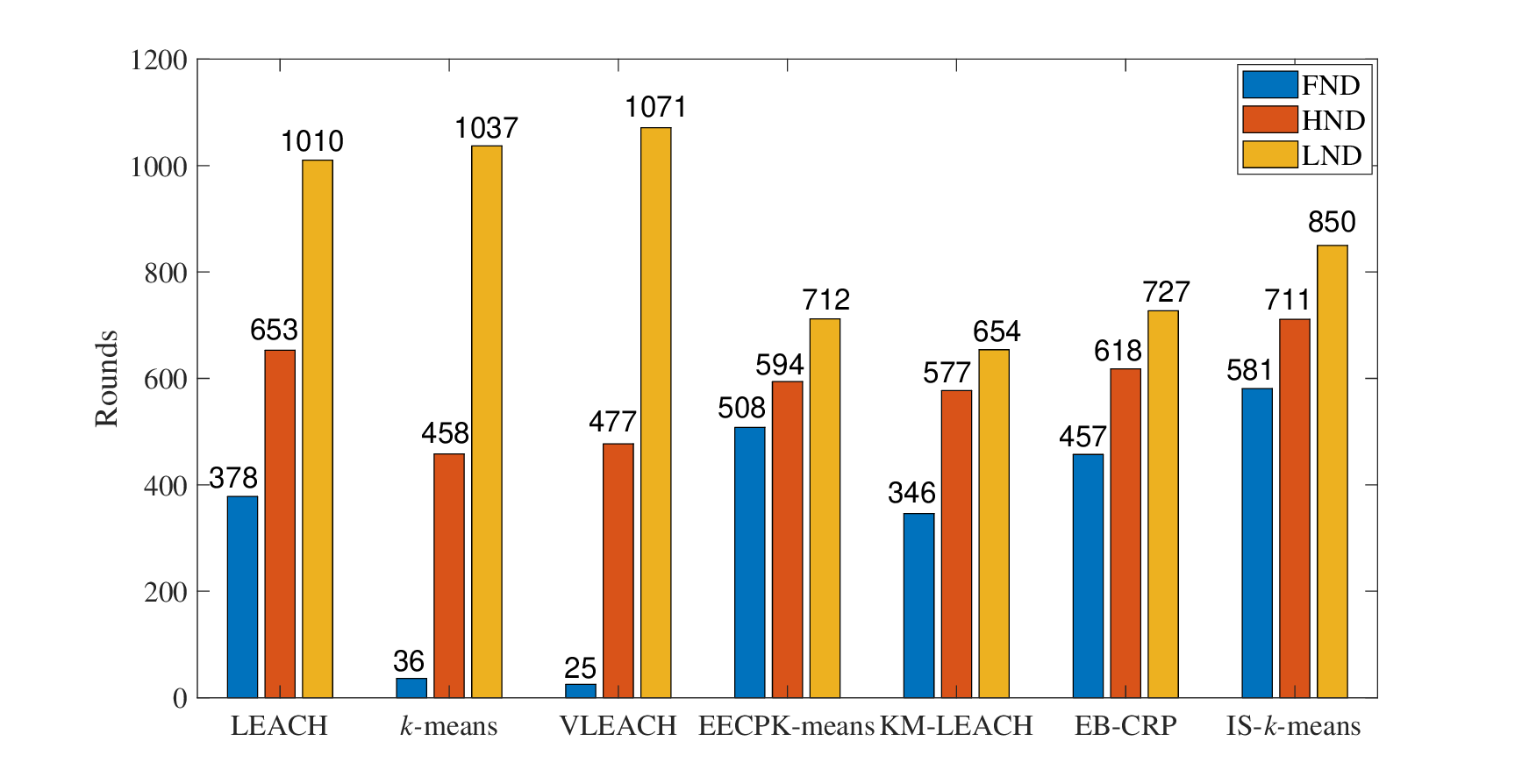}}\hspace{-0.4in}
	\caption{{Comparison of FND, HND, and LND. (a) Scenario 1. (b) Scenario 2.}}
	\label{fig9}
\end{figure*}

\begin{figure*}[!t]
	\centering
	 \hspace{-0.3in}\subfigure[]{\includegraphics[width=0.5\textwidth]{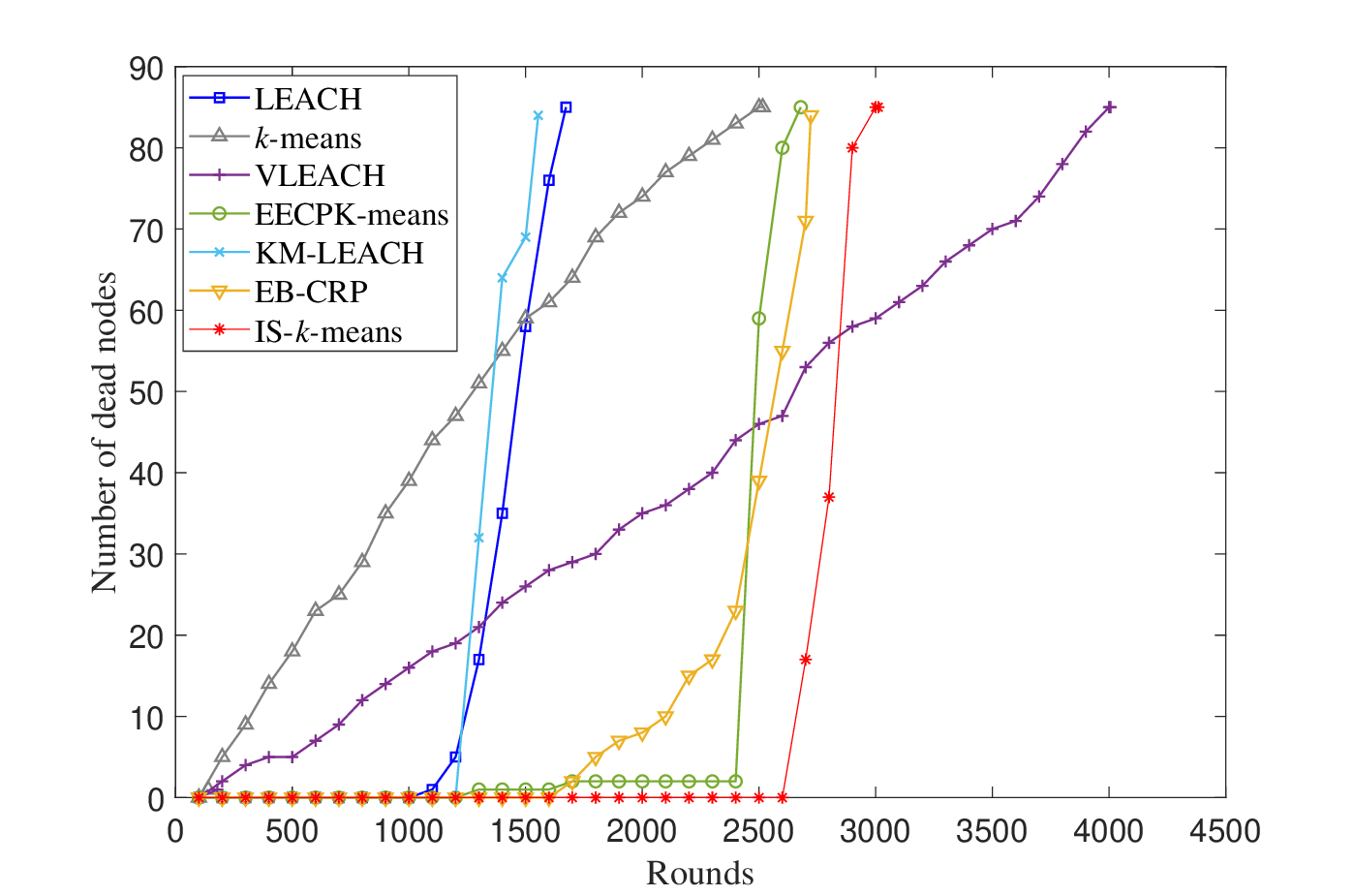}}
	 \hspace{-0.2in}\subfigure[]{\includegraphics[width=0.5\textwidth]{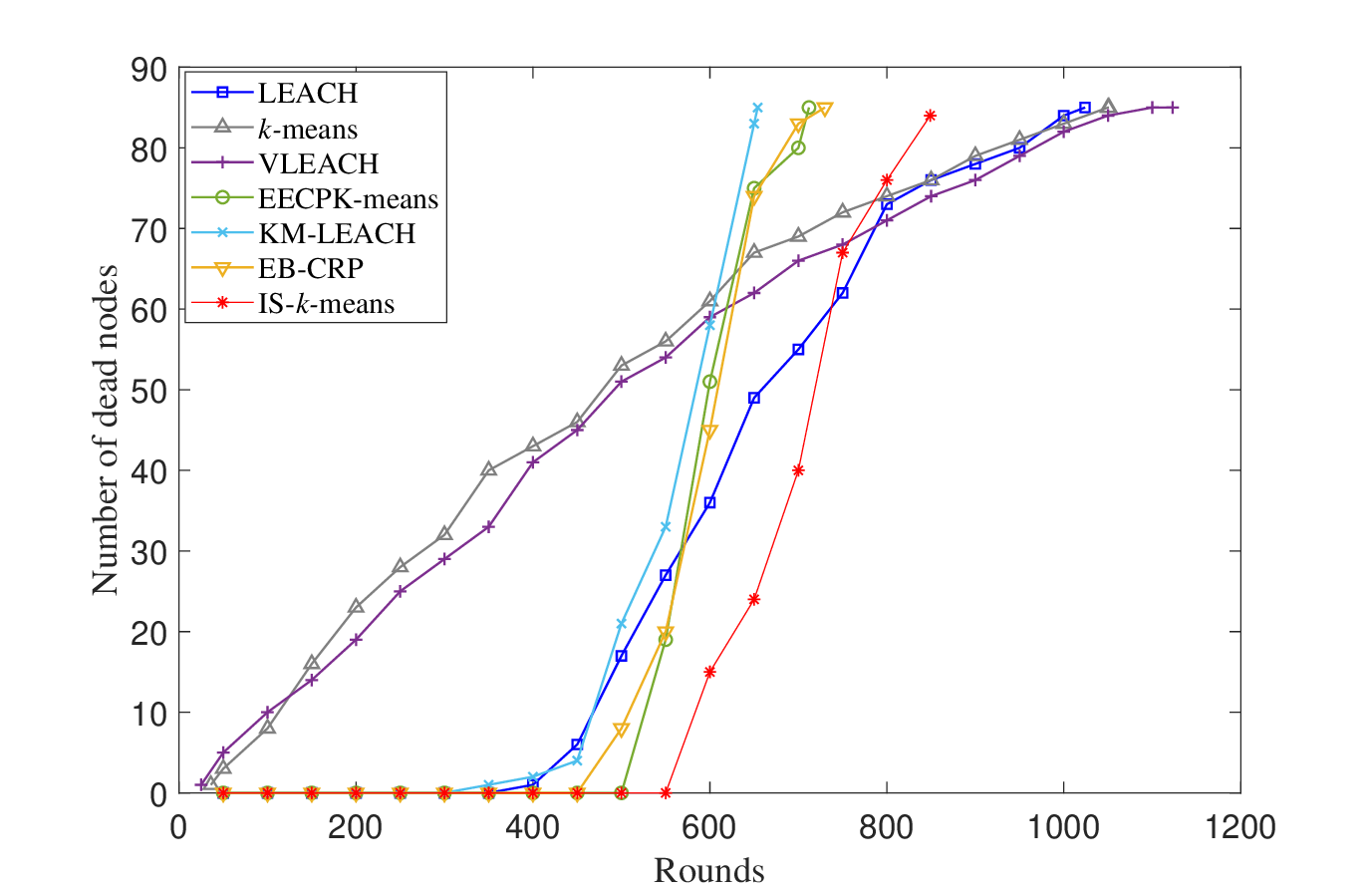}}\hspace{-0.4in}
	\caption{{Comparison of network lifetime of LEACH, \textit{k}-means, VLEACH, EECPK-means, KM-LEACH, EB-CRP, and IS-\textit{k}-means. (a) Scenario 1. (b) Scenario 2.}}
	\label{fig10}
\end{figure*}

 We assume there are 100 sensor nodes that are randomly distributed in both scenario 1 and scenario 2. The obtained results are the averages of 20 independent experiments. The authors in \cite{b10} found the optimum number of clusters to be between 3 and 5 for 100-node network in LEACH. Thus, in  scenario 1, we set 4 as the initial number of clusters in LEACH. For the other six algorithms, we use CFSFDP and KDE to determine the number of clusters in order to ensure the same number of clusters for each algorithm. The initial number of clusters found from the CFSFDP and KDE algorithms is 4 in  scenario 1. In  scenario 2, all algorithms are set with the same number of clusters 6 that is determined by CFSFDP and KDE. We assume that the death of 85\% nodes means all nodes are dead.

Fig. \ref{fig9} shows the first node death (FND), half of nodes death (HND), and the last node death (LND) for these seven algorithms when the number of nodes is 100. If an algorithm can balance energy well, the first node death will be very late. In Fig. \ref{fig9} (a), the average number of rounds of FND in $k$-means is 88, which is much earlier than 970 in LEACH and 2627 in IS-$k$-means, and the average LND happens later when compared to LEACH and the IS-$k$-means algorithms. Thus, it is obvious that the energy consumption of $k$-means is unbalanced. Although VLEACH uses the vice CH scheme in each cluster to extend the network lifetime, it exhibits a poor performance in balancing energy consumption because its FND is 78 and LND is 3979, as shown in Fig. \ref{fig9} (a). EECPK-means improves the selection of initial cluster centers of the $k$-means algorithm by using the midpoint algorithm. It outperforms LEACH and KM-LEACH in both balancing energy consumption and extending network lifetime. For the EB-CRP algorithm, its FND is about 1.7 times that of LEACH, 19 times that of $k$-means, 21 times that of VLEACH, 1.5 times that of EECPK-means, and 1.5 times that of KM-LEACH, which demonstrates that the EB-CRP algorithm can postpone the death of the first node when compared with the other five algorithms. In addition, the HND of EB-CRP is 2579, which is larger than 1462 in LEACH, 1244 in $k$-means, 2458 in EECPK-means, and 1339 in KM-LEACH. This result means that the EB-CRP algorithm can delay the death of the first 50\% of nodes as compared to LEACH, $k$-means, EECPK-means, KM-LEACH. Thus, the EB-CRP shows a good performance in balancing the energy consumption of the nodes and increasing the network lifetime. In view of Fig. \ref{fig9} (a), our proposed IS-$k$-means algorithm can effectively postpone the FND, HND and LND. The average FND of IS-$k$-means is 2627, which is around 2.7 times that of LEACH, 30 times that of $k$-means, 34 times that of VLEACH, 2.3 times that of KM-LEACH, 2.4 times that of EECPK-means, and 1.5 times that of EB-CRP.
Instead of using a fixed number of communication rounds during each steady-phase, like in the EB-CRP algorithm, the communication rounds in our proposed IS-$k$-means algorithm are determined by the residual energy of CHs. If the residual energy of any CH is below the threshold, the algorithm will stop the current steady-phase and trigger re-clustering, which can avoid CHs to die earlier than  EB-CRP. Thus, the IS-$k$-means algorithm can keep all nodes in the network alive in most rounds. The average HND of the IS-$k$-means is also around 2 times among LEACH, $k$-means, and KM-LEACH.

\begin{figure*}[!t]
  \centering
    \subfigure[]{\includegraphics[width=3.6in]{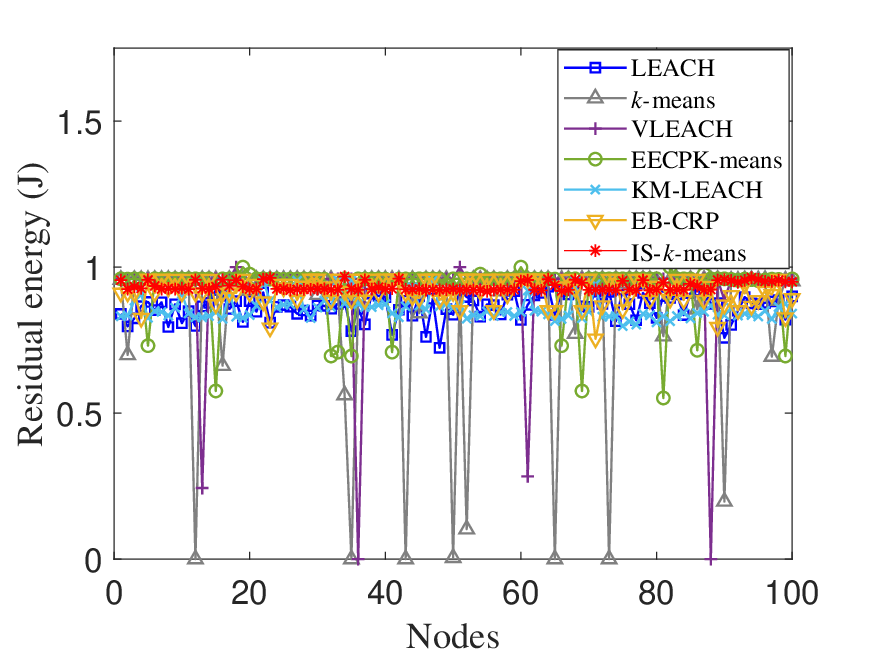}
			\label{Residual energy after 200 rounds}}\hspace{-0.2in}
	\subfigure[]{\includegraphics[width=3.6in]{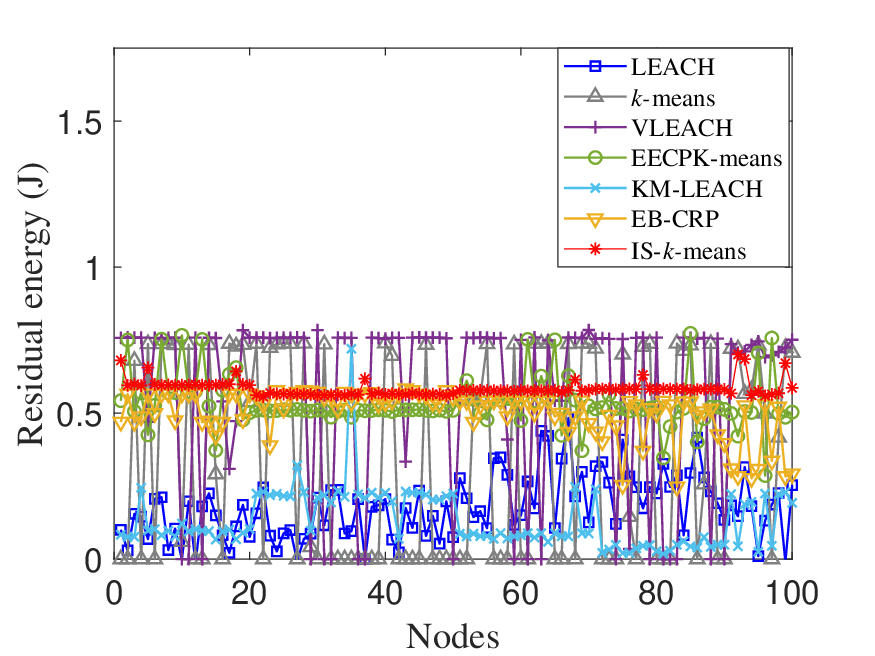}
			\label{Residual energy after 1400 rounds}} \\

    \subfigure[]{\includegraphics[width=3.6in]{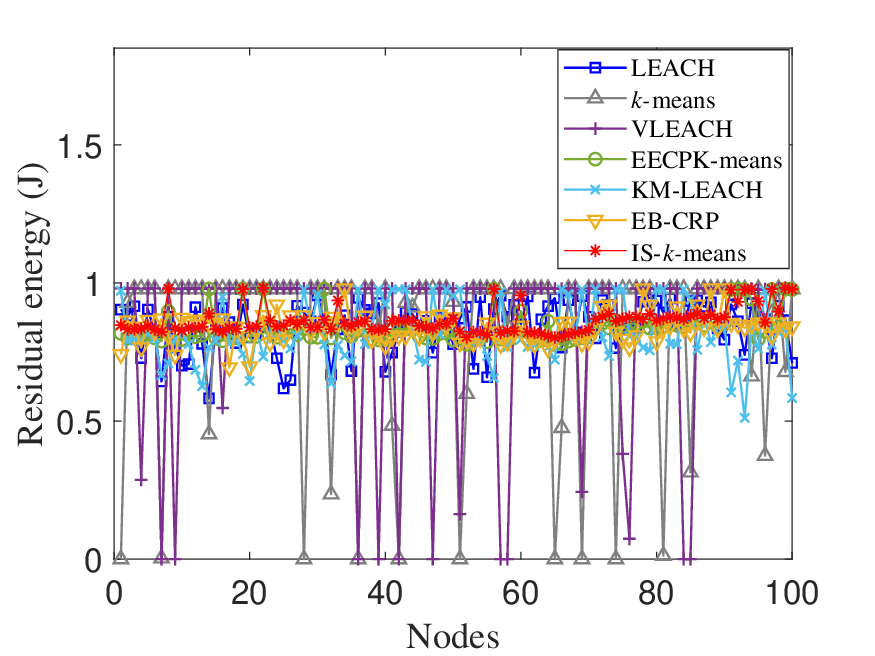}
			\label{}}\hspace{-0.2in}
	\subfigure[]{\includegraphics[width=3.6in]{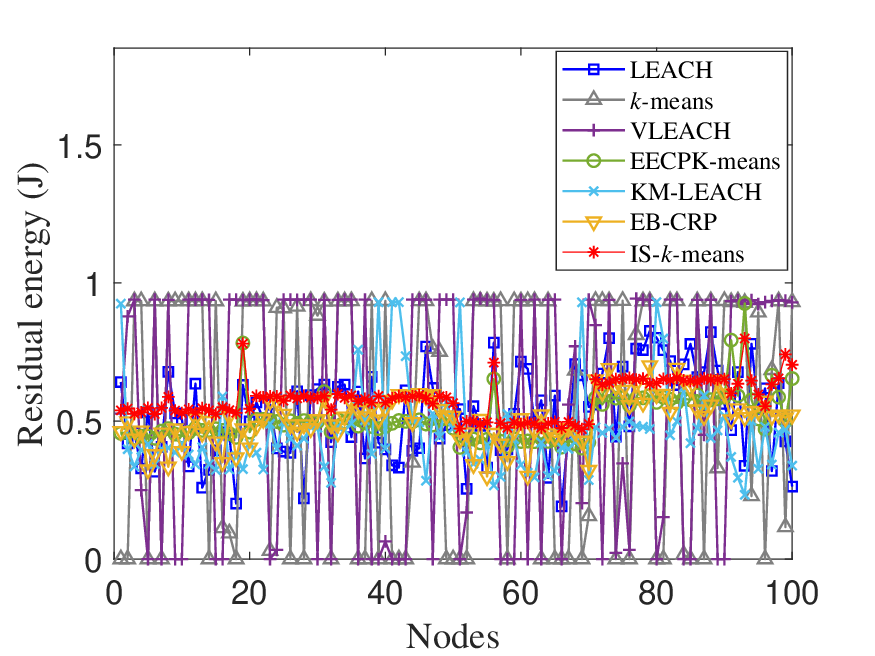}
			\label{}} \\
     \caption{Comparison of residual energy curve. (a) Residual energy after 400 rounds in scenario 1. (b) Residual energy after 1000 rounds in scenario 1. (c) Residual energy after 100 rounds in scenario 2. (d) Residual energy after 300 rounds in scenario 2.}
	\label{fig12}
\end{figure*}

In Fig. \ref{fig9} (b), the average FND, HND, and LND of all algorithms are decreased. This is because extending the network size will increase the communication distance of the nodes which leads to an increase in the energy consumption. The VLEACH and $k$-means algorithms still show a very poor outcome in balancing the energy consumption, a consequence of having small FND and large LND. However, EECPK-means and EB-CRP have relatively large values of the FND and HND, which means most nodes in these two algorithms live longer when compared with $k$-means, VLEACH, and KM-LEACH. In addition, it is evident that our proposed IS-$k$-means algorithm has the best results in postponing the FND, HND, and LND as compared with the other six algorithms.

\begin{table*}[!t]
	\centering
	\caption{Comparison of energy variance of different rounds}
	\label{table3}
	\begin{tabular}{c|c|c|c|c|c|c|c|c}
		\Xhline{1pt}
		 \multicolumn{2}{c|}{} & LEACH & $k$-means & VLEACH & EECPK-means & KM-LEACH & EB-CRP & IS-$k$-means\\ \hline
	    \multirow{7}{*}{{Scenario 1}}	& {200 rounds} & {0.0018} & {0.0374} & {0.0271} & {0.0127} & {0.0013} & {0.0019} & {0.0002}\\
		\cline{2-9}
		& {400 rounds} & {0.0039} & {0.0768} & {0.0496} & {0.0264} & {0.0025} & {0.0024} & {0.0004}
		\\
		\cline{2-9}
		& {600 rounds} & {0.0074} & {0.1161} & {0.0775} & {0.0396} & {0.0085} & {0.0026} & {0.0004}\\
		\cline{2-9}
		& {800 rounds} & {0.0108} & {0.1163} & {0.0756} & {0.0432} & {0.0098} & {0.0038} & {0.0005}\\
		\cline{2-9}
		& {1000 rounds} & {0.0168} & {0.1148} & {0.0882} & {0.0498} & {0.0094} & {0.0044} & {0.0008}\\
		\cline{2-9}
		& {1200 rounds} & {0.0172} & {0.1016} & {0.0904} & {0.0515} & {0.0055} & {0.0067} & {0.0007}\\
		\cline{2-9}
	    & {1400 rounds} & {0.0095} & {0.0988} & {0.0901} & {0.0496} & {0.0024} & {0.0072} & {0.0009} \\
	    \Xhline{1pt}
	    \multirow{6}{*}{{Scenario 2}}	& {100 rounds} & {0.0081} & {0.0983} & {0.1070} & {0.0031} & {0.0112} & {0.0028} & {0.0022} \\
		\cline{2-9}
		& {200 rounds} & {0.0131} & {0.1642} & {0.1645} & {0.0058} & {0.0186} & {0.0052} & {0.0045}
		\\
		\cline{2-9}
		& {300 rounds} & {0.0248} & {0.1921} & {0.1899} & {0.0073} & {0.0258} & {0.0073} & {0.0046}\\
		\cline{2-9}
		& {400 rounds} & {0.0375} & {0.1806} & {0.1963} & {0.0105} & {0.0392}  & {0.0094} & {0.0080}\\
		\cline{2-9}
		& {500 rounds} & {0.0388} & {0.1713} & {0.1821} & {0.0135} & {0.0312} & {0.0122} & {0.0110}\\
		\cline{2-9}
		& {600 rounds} & {0.0357} & {0.1491} & {0.1647} & {0.0167} & {0.0181} & {0.0168} & {0.0141} \\
	    \Xhline{1pt}
	\end{tabular}
\end{table*}

Fig.~\ref{fig10} shows the network lifetime comparison of our proposed IS-$k$-means algorithm and the other six algorithms.
As can be seen from Fig.~\ref{fig10} (a), the network lifetime curves of KM-LEACH, LEACH, EECPK-means, and the proposed IS-$k$-means algorithms are approximately vertical. This means that, in these algorithms, the majority of nodes die approximately after the same number of rounds. Furthermore, one can see that the proposed IS-$k$-means algorithm outperforms KM-LEACH, LEACH, and EECPK-means algorithms in terms of the energy consumption equilibrium. The results in Fig. \ref{fig10} (a) also show that VLEACH  has a longer network lifetime than our proposed IS-$k$-means algorithm. This is reasonable since the objective of VLEACH is to extend the network lifetime, whereas our proposed IS-$k$-means algorithm aims to balance the energy consumption in the network. As a result, some nodes die very early and others die very late in VLEACH, which likely results in the inability to collect sensing data from certain areas where some nodes are dead. In Fig. \ref{fig10} (b), although none of the algorithms shows a nearly vertical curve, like in Fig. \ref{fig10} (a), our proposed algorithm still outperforms the other six algorithms in balancing the energy consumption and prolonging the network lifetime.

\subsection{Energy Variance}

Fig. \ref{fig12} compares the average residual energy of all 100 nodes in WSNs among {the seven} algorithms after different rounds {in two scenarios}. It is found that the residual energy curve of all nodes in the IS-$k$-means algorithm is smoother than that of the other {six} algorithms. This result demonstrates that the IS-$k$-means algorithm is good at balancing the energy consumption of all nodes in WSNs. For the purpose of estimating performance of the proposed algorithm, we introduce a new parameter called energy variance (EV), which is expressed as
\begin{equation}
   \text{EV} = \frac{\sum_{i=1}^{n}\left(E_i\left(r\right) - \overline{E}\right)^2}{n},
\end{equation}
where $\overline{E}$ is the average energy of all nodes. Table \ref{table3} clearly reveals that {EB-CRP has relatively smaller variances than LEACH, $k$-means, VLEACH, KM-LEACH, and EECPK-means in different rounds. In addition, our proposed IS-$k$-means algorithm achieves the smallest variances among seven algorithms}, which demonstrates that the IS-$k$-means can keep the residual energy of 100 nodes to be the most uniform in WSNs.

{It is worthy to mention that the EB-CRP algorithm shows better performance in extending the network lifetime for WSNs with large network sizes\cite{R. Yarinezhad and S. N. Hashemi}, while the proposed algorithm has good performance in balancing the energy consumption and extending the network lifetime for smaller network sizes. We briefly summarize the reasons why the proposed algorithm performs better than the other six algorithms for WSNs of smaller sizes. First, optimizing the initial cluster centers of the soft $k$-means algorithm and reassigning nodes can better balance the number of nodes in different clusters to form good clustering results. Second, our algorithm selects nodes with more residual energy as the CHs, which can prevent the CHs from dying too early and support a high number of communication rounds. Third, the multi-CHs scheme of the proposed IS-$k$-means can reduce the communication energy consumption in the set-up phase caused by re-clustering because it reduces the number of re-clustering. Thus, all sensors can save energy to maintain more communication rounds in the steady phase, which extends the network lifetime. However, for the EB-CRP algorithm, it only chooses one CH in each cluster, which may cause all nodes to re-cluster frequently because CHs may quickly exhaust their energy. Fourth, instead of using a fixed number of communication rounds during each steady-phase, like in EB-CRP, the communication rounds in our proposed IS-$k$-means algorithm are determined by the residual energy of CHs. If the residual energy of any CH is below the threshold, the algorithm will stop the current steady-phase and trigger re-clustering, which can avoid CHs to die earlier than  the EB-CRP algorithm.}


\section{Conclusions}
In this paper, we proposed an energy balanced IS-$k$-means algorithm based on the soft $k$-means for WSNs. The proposed algorithm improves the selection of initial cluster centers by using CFSFDP and KDE algorithms. In order to balance the number of nodes per cluster, the proposed algorithm reassigns nodes at the edge of different clusters to a low-density cluster according to the nodes' membership probabilities. Furthermore, multi-CHs scheme was used in the selection of final CHs, which can effectively balance the traffic load of CHs, reduces the number of re-clustering and saves communication cost in the set-up phase. In order to show the advantages of the IS-$k$-means in balancing energy consumption, we compared it with LEACH, $k$-means, VLEACH, EECPK-means, KM-LEACH, {and EB-CRP. In scenario 1,} simulation results demonstrated that the proposed IS-$k$-means algorithm postponed the FND by {2.7 times, 34 times}, 2.3 times, 2.4 times, 30 times, {and 1.5 times} when compared to LEACH, VLEACH, KM-LEACH, EECPK-means, $k$-means, {and EB-CRP} on average, respectively. The HND of the IS-$k$-means algorithm also was delayed by 2 times when compared to LEACH, $k$-means, and KM-LEACH. {In addition, the IS-$k$-means algorithm achieved an excellent result in postponing the FND and HND in scenario 2 as compared with other mentioned algorithms.} The IS-$k$-means algorithm also extended network lifetime {in both scenarios} as compared to KM-LEACH, {EECPK-means, and EB-CRP}. Furthermore, the proposed algorithm also yields smoother average remaining energy curves of all nodes in different rounds and smaller average energy variances. Hence, the proposed IS-$k$-means algorithm is promising in balancing energy consumption in WSNs. {In a future work, we plan to design an energy-efficient multi-hop routing algorithm to extend the IS-$k$-means algorithm to large-scale WSNs.}


\vspace{-10mm}

\end{document}